\documentclass[preprint,12pt]{elsarticle}
\usepackage{color}
\usepackage{xparse}
\usepackage{etoolbox}
\definecolor{red}{rgb}{1,0,0}
\definecolor{green}{rgb}{0,1,0}
\definecolor{blue}{rgb}{0,0,1}

\usepackage{tikz}

\newrobustcmd*{\mysquare}[1]{\tikz{\filldraw[draw=#1,fill=#1] (0,0)
rectangle (0.2cm,0.2cm);}}

\newrobustcmd*{\myemptysquare}[1]{\tikz{\draw[draw=#1,line width=0.35mm] (0,0)
rectangle (0.2cm,0.2cm);}}

\newrobustcmd*{\mycircle}[1]{\tikz{\filldraw[draw=#1,fill=#1] (0,0) circle [radius=0.1cm];}}

\newrobustcmd*{\myemptycircle}[1]{\tikz{\draw[draw=#1,line width=0.35mm] (0,0) circle [radius=0.1cm];}}

\newrobustcmd*{\mydowntriangle}[1]{\tikz{\filldraw[draw=#1,fill=#1] (0,0.2cm) --
(0.2cm,0.2cm) -- (0.1cm,0cm);}}

\newrobustcmd*{\myemptydowntriangle}[1]{\tikz{\draw[draw=#1,line width=0.35mm] (0,0.2cm) -- (0.2cm,0.2cm) -- (0.1cm,0cm)-- (0, 0.2cm);}}

\usepackage{graphicx}
\usepackage{amssymb}
\usepackage{amsmath}
\usepackage{bm}
\usepackage{subcaption}
\usepackage{multirow}
\usepackage{array}
\usepackage[thinlines]{easytable}
\usepackage{siunitx}

\journal{International Journal of Multiphase Flows}

\begin{document}

\begin{frontmatter}



\title{Direct numerical simulation of two-phase pipe flow: influence of the domain length on the flow regime}

\author[1]{Carlos Plana}
\author[2]{Baofang Song}
\author[1]{Marc Avila\corref{cor1}}
\ead{marc.avila@zarm.uni-bremen.de}
\cortext[cor1]{Corresponding author}
\address[1]{University of Bremen, Center of Applied Space Technology and Microgravity (ZARM), 28359 Bremen, Germany}
\address[2]{Tianjin University, Center for Applied Mathematics, Tianjin 300072, China}


\begin{abstract}


We investigate the dynamics of a kerosene-water mixture in a vertical pipe flow by solving the Cahn--Hilliard--Navier--Stokes equations. We compute the linear stability of laminar core-annular flow  in a vertical pipe and find that it is highly unstable. By performing direct numerical simulations initialized with a slightly perturbed core-annular flow, we show that the system transitions to turbulence and finally relaxes into a turbulent slug flow regime provided that the pipe is sufficiently long. This configuration presents mild turbulence and large scale three-dimensional recirculation patterns. Our work highlights the need for applying nonlinear-dynamics approaches and carefully selecting the domain length to investigate the patterns observed in two-phase pipe flows and demonstrates the capabilities of phase-field methods to reliably simulate flows under realistic experimental conditions.

\end{abstract}


\end{frontmatter}


\section{Introduction} \label{introduction}

The flow of two immiscible fluids through a circular pipe is of interest for many industrial situations \cite{Brauner2003}. The great complexity of multiphase systems allows for a large variety of configurations with very different properties, including emulsions, drops and bubbles, slugs, froths or stratified and core-annular flows \cite{Ibarra2014}. The emergence of a particular regime depends on the flow conditions, such as the volume fluxes of the components, the pipe orientation and the fluid properties, for a total of eight independent dimensionless parameters (see \S\ref{sec:problem} for their definitions). Many laboratory experiments have been conducted to determine the regime dependence on the parameters, but the obtained flow-regime maps are usually valid for a small range of operating conditions and fluid properties, (see for example \cite{Govier1961,Charles1961,Shi2017,Arney1993} for heavy oils and \cite{Angeli2000,Jana2006,Ghosh2012,Wegnan2006} for lighter oils).
Likewise, numerous numerical simulations have been carried out to better understand the dynamics of multiphase flows in simple geometries. A lot of attention has been given to the bubbly regime in turbulent flow, which has been mainly studied in the channel flow geometry, using phase-field methods \cite{Scarbolo2015}, VoF \cite{Cifani2018}, level-set \cite{Bolotnov2011} or front-tracking methods \cite{Lu2006}, but also in homogeneous isotropic turbulence \cite{Trontin2010} and homogeneous shear turbulence  \cite{Rosti2019}. Less chaotic regimes, such as Taylor bubbles \cite{Zimmer2018}, wavy core-annular flow \cite{Li1999} or lubricated channel flow \cite{Roccon2019} have also been investigated.

In the transport of oils in pipes, it is desirable to deliver a target flow rate, whilst minimizing the driving pressure gradient. Core-annular flows (CAF), in which a fluid of high viscosity (core) is surrounded by a fluid of lower viscosity (annulus), exhibit low pressure losses and are thus ideally suited for the transport of very viscous fluids, such as heavy oils \cite{Joseph1997}. For vertical pipes, the governing equations admit an analytical laminar solution (hereafter perfect core-annular flow, PCAF), in which the velocity profile is unidirectional and parabolic in the annulus and the core \cite{Joseph1997}. However, this solution is unstable in most operating regimes, which means that even the slightest disturbances can destabilize it and lead to a different flow configuration. As a result PCAF has never been reported in laboratory experiments. Although this may suggest that the PCAF is irrelevant for the dynamics of vertical two-phase flows, many of the flow patterns that have been reported in experiments of very viscous oils in water are qualitatively similar to PCAF~\cite{Bai1992}. Furthermore, linear stability analyses of the basic flow are capable of predicting some properties of the experimental flow patterns  (e.g.\ wavy CAF), highlighting the importance of PCAF \cite{Bai1992,Hu1989}.  Bai \emph{et al.}~\cite{Bai1992} provided a classification of flow regimes for vertical two-phase pipe flows with a heavy oil-water mixture and showed that wavy CAF are stable in wide parameter regimes. Jana \emph{et al.}~\cite{Jana2006} and Ghosh \emph{et al.}~\cite{Ghosh2012} obtained similar regime maps for the water-kerosene mixture in the upflow and downflow regimes.

In this paper, we investigate numerically the flow of kerosene and water in a vertical pipe. We show that the basic flow consisting of a kerosene core surrounded by a water annulus is highly unstable. For a specific parameter set, we perform interface-resolving direct numerical simulations of the Cahn--Hilliard--Navier--Stokes equations \cite{Jacqmin1999}. We find that the core-annular flow transitions to turbulence and ultimately settles to a saturated regime that is determined by the computational pipe length. For  sufficiently long pipes, the system evolves into a slug flow regime, with mild turbulence and large scale three-dimensional recirculation patterns inside the slugs. 
\section{Problem specification}
\label{sec:problem}

We consider the two-phase pipe flow of water (density $\rho_w=1000$ kg/m$^3$/s, dynamic viscosity $\mu_w=0.001$ kg/m/s) and kerosene ($\rho_o=790$ kg/m$^3$/s, $\mu_o=0.0016$ kg/m/s) in a vertical pipe of diameter $D=0.01$ m, similar to experiments \cite{Ghosh2012, Bai1992}. The surface tension is $\sigma =0.048$ J/m$^2$. In laboratory experiments, the oil is usually injected concentrically at a prescribed flow rate $Q_o$ in a pipe in which water flows at rate $Q_w$ \cite{Bai1992}. In our numerical method~\cite{Song2019}, we employ periodic boundary conditions in the axial direction. Axially periodic pipes do not suffer from end effects and enable the efficient simulation of fully developed flows in relatively short domains (provided that they are not shorter than the typical axial wavelength of the dominant flow pattern). In such simulations, the flow rates of the two phases cannot be chosen independently of each other. Instead, the volumes of the two phases ($V_w$ and $V_o$, with total volume $V=V_o+V_w$) and the total flow rate $Q=Q_o+Q_w$ are imposed. The hold-up ratio~\cite{Govier1961}
\begin{equation} \label{eq:hold-up-ratio}
	h = \frac{\hat{Q}}{\hat{V}},
\end{equation}
where $\hat{Q}= Q_o/Q_w$ is the volume flow ratio and $\hat{V} = V_o/V_w$ the volume ratio, is often used to characterize the flow regimes and enables a direct comparison between experiments and numerical simulations. In this paper, we set $\hat{V}=1.78$, which is in the same range of those reported by Bai \emph{et al.}~\cite{Bai1992}. This value keeps the mass leakage manageable \cite{Yue2007,Song2019}, with the total change of $\hat{V} < 6\%$, and therefore no methods to minimize the leakage \cite{Soligo2019} were implemented here. As the simulation evolves, the change in $h$ reflects changes in the ratio of volume flows, $\hat{Q}$,  as the flow pattern evolves. This is in contrast to experiments, where $\hat{Q}$ is fixed and $\hat{V}$ is an outcome of the experiment.

In general, two-phase pipe flow is governed by eight independent dimensionless parameters
\begin{equation}\label{eq:dimless_params}
Re=\dfrac{\rho_m U D}{\mu_m},\,  We=\dfrac{\rho_m U^2 D}{\sigma},\, Fr=\dfrac{U^2}{D g}, \,
\hat\rho=\dfrac{\rho_w}{\rho_o},\, \hat \mu=\dfrac{\mu_w}{\mu_o}, \, \hat V=\dfrac{V_o}{V_w}.
\end{equation}
The Reynolds $Re$, Weber $We$ and Froude $Fr$ numbers are defined with the mean speed $U=Q/A$ and the volume-averaged density and dynamic viscosity 
\begin{equation}
\rho_m=\dfrac{\rho_wV_w+\rho_oV_o}{V},\quad \mu_m=\dfrac{\mu_wV_w+\mu_oV_o}{V}.
\end{equation}
In our simulations, we fixed the mean speed  $U=0.376$m/s and $g=9.81$m/s$^2$. The two remaining dimensionless parameters of two-phase pipe flow are the pipe inclination, $\gamma$, and the interface-wall contact angle, $\theta_w$. We consider upward flow in a vertical pipe $\gamma=\pi/2$. Depending on the pipe material and surface treatment, it can present a range of interface-wall contact angles, from hydrophilic ($\theta_w < \pi/2$) to hydrophobic ($\theta_w > \pi/2$) \cite{daSilva2006}. We selected a neutral pipe wall, with $\theta_w=\pi/2$, for which there is no preference for either water or kerosene to wet the wall. The corresponding values of the dimensionless parameters are given in table~\ref{tab:dimless_params}.

\begin{table}[h!]
  \begin{center}
    \begin{tabular}{c c c c c c c c}
      \hline
       Re & We & Fr & $\hat{V}$& $\hat{\rho}$ & $\hat{\mu} $ & $\gamma$ & $\theta_w$ \\ [0.2ex]
      \hline
       $2268$  & $25.44$ &  $1.44$ & $1.78$  & $1.266$  & $0.625$ & $\pi/2$ & $\pi/2$\\
      \hline
    \end{tabular}
    \caption{Dimensionless parameters of the vertical, upward two-phase pipe flow of water and kerosene investigated in this paper. For a definition of the parameters see eq.~\eqref{eq:dimless_params} and surrounding text.} \label{tab:dimless_params}
  \end{center}
\end{table}

\subsection{Governing equations}

In order to compute two-phase pipe flows, we solved the Cahn--Hilliard--Navier--Stokes (CHNS) equations \cite{Jacqmin1999,Anderson1998}, in which the phase variable $C\in[-0.5,0.5]$ denotes the composition of the fluid mixture; $C=-0.5$ corresponds to pure water and $C=0.5$ to pure oil. In the CHNS, $C$ varies smoothly across diffuse interfaces between these two values. All quantities were rendered dimensionless by scaling lengths with the diameter $D$, velocities with the mean speed $U$, time with the advective time unit $D/U$ and pressure with the viscous pressure scale $\mu_m U/D$. The dimensionless (CHNS) equations \cite{Jacqmin1999,Anderson1998} read
\begin{align}
\nabla \cdot \boldsymbol u &= 0,
\\
\begin{split}
\label{dimlessN-s}\rho(C) Re \left(\dfrac{\partial\boldsymbol u}{\partial t} +  \boldsymbol u \cdot \nabla \boldsymbol u \right)  &= \\
 - \nabla \tilde{p} +  \nabla \cdot &  \boldsymbol T -  \frac{\sqrt{18}Re}{WeCn}C\nabla \Phi  + Re \left(-f \boldsymbol  e_z +\frac{\rho(C)}{Fr} \boldsymbol g \right),
\end{split}
\\
\dfrac{\partial C}{\partial t} + \boldsymbol u\cdot \nabla C &= \frac{1}{Pe} \nabla^2 \Phi.\label{equ:CH_dimless}
\end{align}
Here $\boldsymbol u$ is the fluid velocity and $\tilde p$ the generalized pressure, $\tilde p=p - C\Phi + \beta\Psi - \alpha/2|\nabla C|^2$, which is a combination of the true fluid pressure $p$ and the potential terms of the surface tension force. The viscous stress tensor is $\boldsymbol T=\mu(C)( \nabla \boldsymbol u + \nabla \boldsymbol u^T)$ and $f(t)$ is the (negative) dimensionless pressure gradient necessary to drive the flow at dimensionless speed $1$. In this paper a vertical pipe is simulated, i.e.\ $\boldsymbol{g} =-\boldsymbol{e_z}$. The viscosity ratio $\hat \mu$ and the density ratio $\hat \rho$ enter the equations through the dimensionless viscosity $\mu(C)$ and density $\rho(C)$, which are calculated as a linear combination of the single-phase properties. The motion-causing component of the surface tension appears in the right-hand-side of the momentum equation~\eqref{dimlessN-s} and has the form $C\nabla \Phi$, where 
\begin{equation}
\Phi = \frac{\partial \Psi}{\partial C} - Cn^2\nabla^2C,
\end{equation}
is the chemical potential of the liquid mixture. Its first term represents the bulk free energy, modeled with the double well potential $\Psi(C) = (C+0.5)^2(C-0.5)^2$ and its second term the interface free energy. For a more detailed derivation of the method, see e.g. \cite{Jacqmin1999}.

The Cahn and the Peclet numbers defined as
\begin{equation}\label{chns_params}
Cn=\dfrac{\sqrt{\alpha/\beta}}{D},\quad
Pe=\dfrac{U D}{\kappa\beta},
\end{equation}
are model parameters that represent the dimensionless interface width ($\epsilon \approx 4.16 Cn$) and the dimensionless inverse of the interface mobility, respectively. These parameters control how the sharp-interface limit is approached \cite{Jacqmin1999}. In particular
\begin{equation}\label{equ:Pen_relation}
Pe\propto Cn^{-1},
\end{equation}
as $Cn \rightarrow 0$ is required in order to recover the correct interface dynamics \cite{Magaletti2014}. Magaletti \emph{et al.}~\cite{Magaletti2014} suggested $Pe = 1/(3Cn)$. In our previous work \cite{Song2019}, we showed that a smaller pre-factor allows for larger time-step sizes without sacrificing accuracy. According to those results, we employed $Pe = 1/(9.27 Cn)$ in all simulations. At the pipe wall, we applied non-slip boundary condition for the velocity and  $\boldsymbol n \cdot \nabla (\nabla^2 C)=0$ and $\boldsymbol n \cdot \nabla C =0$ for the Cahn--Hilliard equation. The former ensures no flux of the phase variable through the wall and the later sets $\theta_w = \pi/2$. This is a simplification of the more general boundary condition for an arbitrary dynamic contact angle \cite{Jacqmin1999},
\begin{equation} \label{eq:general_angle}
	\frac{D C}{D t} = D_w \left( \alpha \frac{\partial C}{\partial n} + \gamma \frac{\partial g}{\partial C} \right),
\end{equation}
 where $D_w$ is the wall diffusion coefficient and $\gamma g(C)$ represents the free energy at the wall as a function of the composition. If we assume $\frac{D C}{D t}=0$  for an equilibrium static contact angle and set $g(C)$ as a constant, meaning that no phase preferentially wets the wall, equation \eqref{eq:general_angle} reduces to  $\boldsymbol n \cdot \nabla C =0$ for $\theta_w=\pi/2$, as assumed in this work.

\subsection{Numerical method}\label{sec:method}

 We solved the CHNS in cylindrical coordinates $(r,\theta,z)$, using the finite-differences method with a 7-point stencil for the radial discretization and the Fourier--Galerkin spectral method for the axial and azimuthal periodic directions. Variables ($\boldsymbol u$, $\tilde{p}$ and $C$) are  written as
\begin{equation}\label{equ:Fourier_expansion}
A(r,\theta,z,t)=\sum_{k=-K}^{K}\sum_{m=-M}^{M}\hat{A}_{k,m}(r,t)e^{i(k_0 kz + m_0m\theta)},
\end{equation}
where $k_0k$ is the axial wavenumber and $m_0m$ the azimuthal wavenumber of the Fourier mode $(k,m)$; $2K$ and $2M$ are the number of modes in the axial and azimuthal directions. The dimensionless pipe length is $L_z=2\pi/k_0$, and $m_0$ is typically set to 1 to recover the complete circumference in azimuthal direction. $\hat{A}_{k,m}$ denotes the complex Fourier coefficient of mode $(k,m)$. We adopted the Crank--Nicolson time integration scheme and the influence matrix method for the treatment of the incompressibility condition described in \cite{Guseva2015}. For the solution of the Cahn--Hilliard equation and the treatment of the variable terms in the Navier--Stokes equation, we followed \cite{Dong2012}. The evaluation of the nonlinear terms was performed using the pseudo-spectral technique (with the $3/2$-rule for de-aliasing, implying 3K and 3M axial and azimuthal points in physical space for the computation of nonlinear terms), which utilizes the fast Fourier transform (FFT) to convert data between physical and spectral spaces. We used the MPI-OpenMP hybrid parallelization strategy of \cite{Shi2015}. The implementation is based on and extends the open-source single-phase code {\tt nspipe} \cite{Lopez2020} and has been extensively validated in \cite{Song2019}, where further details of the method can be found. 

\subsection{Perfect core-annular flow (PCAF)} 

\begin{figure}[!ht]
        \centering
            \centering 
            {{\small }}   
            \includegraphics[scale=1]{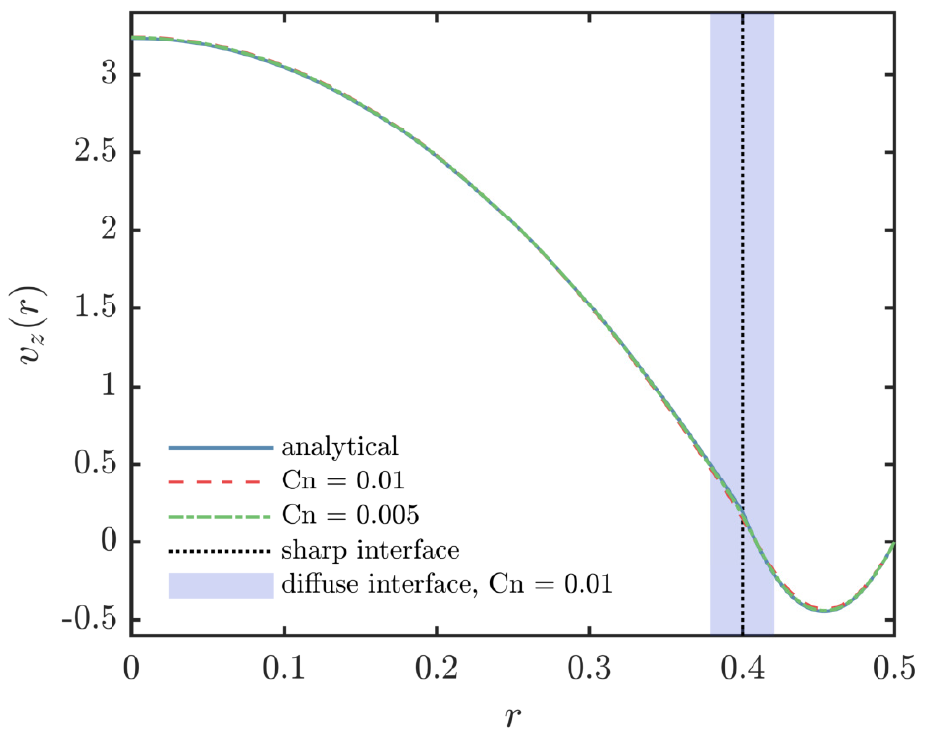}
        \caption[]{\label{fig:LCAF} 
Upward perfect core-annular flow (PCAF) of water (annulus) and kerosene (core). The values of the dimensionless flow parameters are given in table \ref{tab:dimless_params}. Solid line: analytical solution to the Navier--Stokes equations in the sharp-interface limit. Dashed (dashed-dotted) line: corresponding unidirectional numerical solution $u_z(r)$ of the CHNS for $Cn=0.01$ ($Cn=0.005$). 
Vertical dotted line: Sharp interface position ($\eta = 0.4$). Shaded area: diffuse interface of width, $\epsilon=4.16Cn$ with $Cn=0.01$. 
}    
\end{figure}

In the sharp-interface limit, the Navier--Stokes equations (with classical stress balance at the fluid interfaces) admit an analytical solution (PCAF) for the specific case of two-phase flows in vertical pipes \cite{Joseph1997}. This solution is shown as a solid blue line in figure~\ref{fig:LCAF} for the parameters considered in this work, with the sharp interface position ($\eta=0.4$) marked as a vertical dotted line. The corresponding numerical approximation computed by solving the CHNS with $Cn=0.005$ and  $Cn=0.01$, using $n_r=192$ and $n_r=96$ respectively, $K=M=0$ and $\Delta t=5\cdot10^{-4}$ are shown to be in excellent agreement with the analytical one, with relative errors $\epsilon_{||L2||,Cn=0.005}=0.015$ and $\epsilon_{||L2||,Cn=0.01}=0.03$. For the selected parameters, most of the water flows downwards (see the negative fluid velocities close to the pipe wall). This is because the downward pull of gravity overcomes the combined force of the driving pressure gradient and the momentum transfer from the core (kerosene) phase.

\section{Linear stability analysis}
\label{sec:stability}

\begin{figure}[!ht]
\begin{subfigure}[c]{0.43\textwidth}
\includegraphics[scale=1]{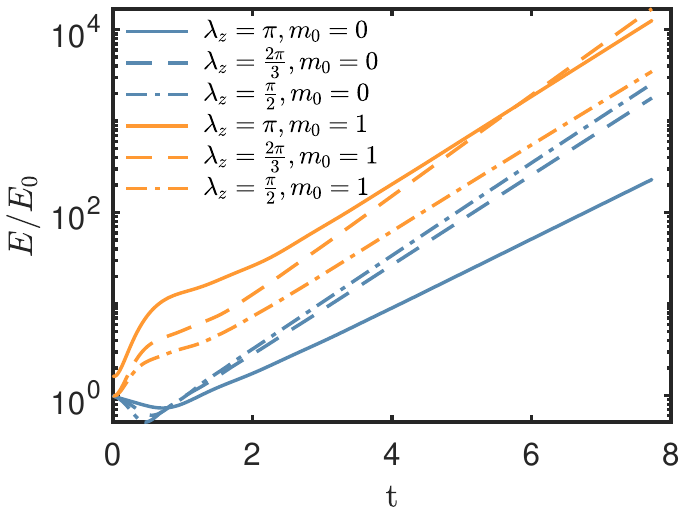}
\subcaption{} \label{fig:laminar_evol}  
\end{subfigure}
\hspace{1cm}
\begin{subfigure}[c]{0.43\textwidth}
\includegraphics[scale = 1]{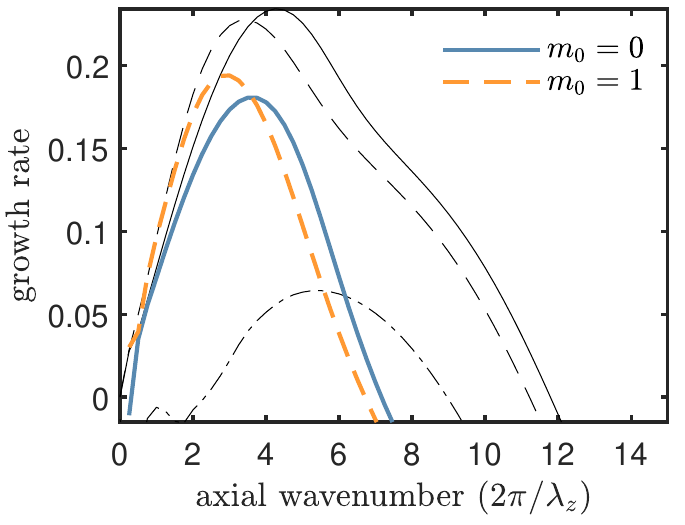}
\subcaption{ } \label{fig:laminar_gr}  
\end{subfigure}
\caption{Linear stability analysis of PCAF. a) Temporal evolution of the most unstable perturbations which fit in pipes of length $L_z=4\pi$;  $Cn=0.01$. The legend indicates the perturbation's axial wavelength ($\lambda_z=2\pi/k_0$) and azimuthal wavenumbers ($m_0$).  b) Linear growth rate of the axisymmetric ($m_0=0$) and non-axisymmetric ($m_0=1$, $2$) modes as a function of the axial wavenumber. Thick (colored) lines correspond to $Cn=0.01$ and thin (gray) lines to $Cn=0.005$. Line types denote the azimuthal wavenumber, with solid/dashed/dash-dotted lines corresponding to $m_0=0/1/2$.}
\end{figure}

In order to estimate the length scale of the eventual structures developing from PCAF, a linear stability analysis was performed with our time-stepping code. Starting with the PCAF velocity profile shown as a dashed line in figure \ref{fig:LCAF}, a small disturbance in the form of a constant, small value of the radial velocity ($v_r = 10^{-10}$) was added to a single Fourier mode with selected axial and azimuthal wavenumbers, $k_0$ and $m_0$ in eq.~\eqref{equ:Fourier_expansion}. The governing equations were integrated in time for this single mode (by setting $K=M=1$), which is equivalent to time-stepping the linearized Navier--Stokes equations, provided that the amplitude of this mode remains sufficiently small. The time step was set to $\Delta t = 5\cdot10^{-4}$. This procedure was repeated by varying the axial wavenumber $k_0$ in small steps and the azimuthal wavenumber $m_0\le 5$ in order to detect the most dangerous perturbation (with largest growth rate). An extensive validation of our method  against a formal linear stability analysis in the sharp-interface limit can be found in \cite{Song2019}. In figure~\ref{fig:laminar_evol}, we show the evolution of the energy of disturbances for selected unstable Fourier modes, which are relevant for the direct numerical simulations shown in the next section. For $t\lesssim 3$, non-axisymmetric disturbances exhibit significant non-modal (algebraic) transient energy growth \cite{Orazzo2014}, before settling into the exponential growth characteristic of a linear instability. By contrast, the axisymmetric disturbances used here do not exhibit non-modal growth and rapidly settle into exponential growth. From the slope of the lines in figure~\ref{fig:laminar_evol}, the leading eigenvalue for each Fourier mode (the growth rate) can be extracted. 
    
\begin{figure}[!ht]
\includegraphics[width=1\linewidth]{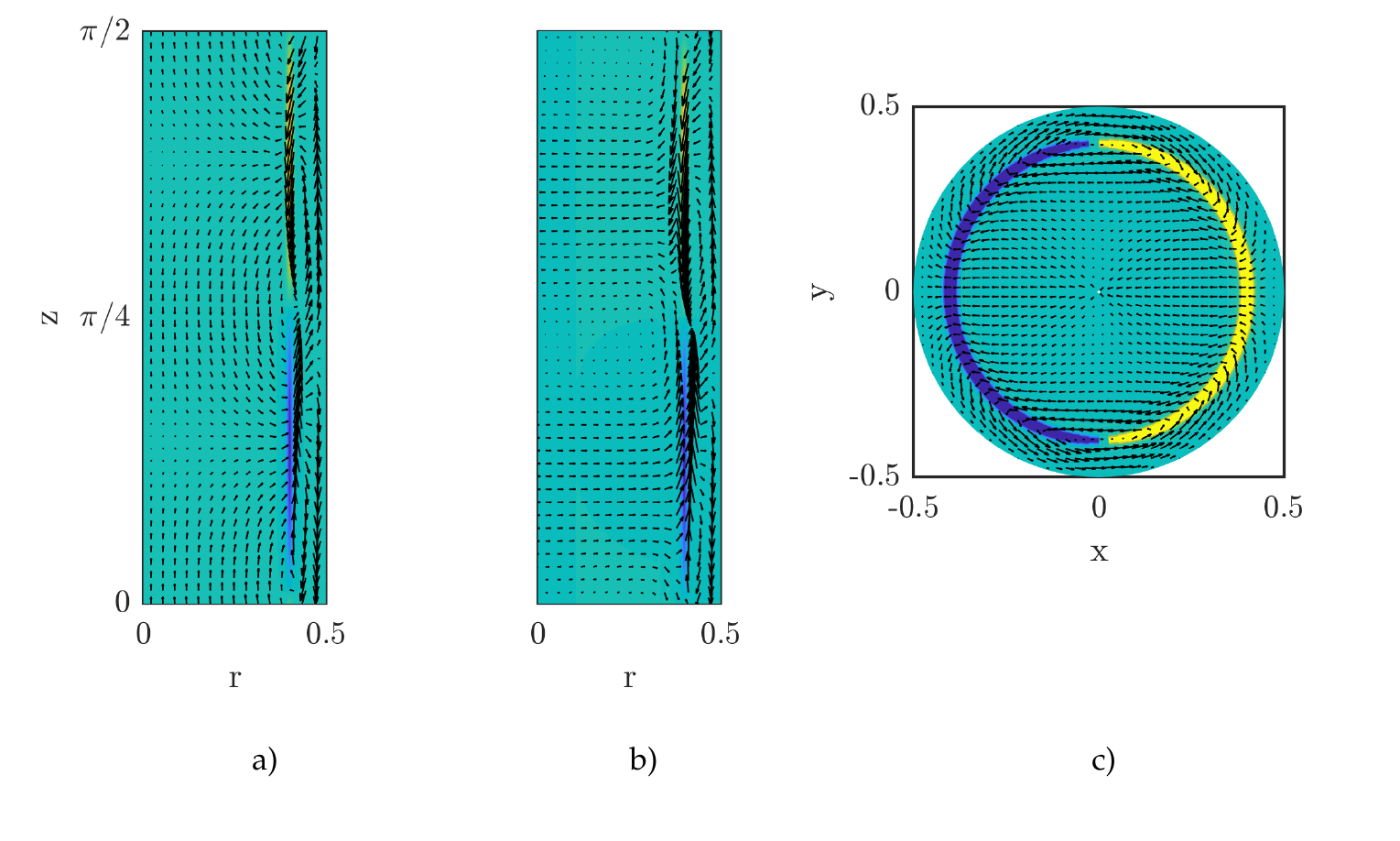}
\caption{In-plane velocity field (vectors) and phase field C (colormap) of the most unstable eigenmodes used in the DNS simulation. a) rz-plane, mode ($\lambda_z=\pi/2$, $m=0$). b) $(r,z)$-plane, mode ($\lambda_z=\pi/2$, $m=1$). c) $(r,\theta)$-plane, mode ($kk_0=4$, $m=1$). }\label{fig:eigenmodes}
\end{figure}
 
In figure~\ref{fig:laminar_gr}, we show the computed growth rate as a function of the (continuous) axial wavenumber for $Cn=0.01$. The non-axisymmetric $m=1$ mode dominates, with a peak at $k_0\approx 2.9$, but the axisymmetric $m=0$ mode exhibits similar growth rate, with a peak at $k_0\approx 3.6$. Perturbations with larger azimuthal wavenumbers $m\ge2$ are stable for all axial wavenumbers. The thin (black) lines in figure~\ref{fig:laminar_gr} depict the same curves but for smaller interface thickness, $Cn=0.005$. At low axial wavenumbers the results agree very well with those for $Cn=0.01$, but progressively deviate as $k_0$ increases, a trend already reported in \cite{Song2019} for core-annular flow of highly viscous oils in water. For $Cn=0.005$, the most unstable perturbation is axisymmetric ($m_0=0$) with axial wavenumber $k_0 \approx 4.2$. The non-axisymmetric mode ($m_0=1$) shows a similar maximum value and position, $k_z\approx3.4$. Shorter azimuthal waves ($m_0=2$) are unstable in a narrow range of axial wavenumbers for $Cn=0.005$ and with  $m\geq3$ are stable throughout. The discrepancies with $Cn=0.01$ are due to the increased interface dissipation at larger $Cn$, which acts much more strongly on larger wavenumbers. While interface dissipation has a relatively large impact in the calculation of the linear growth rate, its influence in the non-linear regime is much lower, as shown in \cite{Song2019} and later in section~\ref{sec:res_test}. 

The velocity field and interface deformation of the leading eigenmodes with $k_0=4$ and $m_0=0,1$ is shown in figure~\ref{fig:eigenmodes} for $Cn=0.005$. The largest fluid velocities are found close to the interface between the two fluids and suggest a Kelvin--Helmholtz instability type associated to the shear at the interface.
 
 
\section{Direct numerical simulation}
\label{sec:3D}

We investigated the nonlinear dynamics of the system by performing three-dimensional, fully nonlinear simulations (DNS) of the CHNS.  We set $Cn=0.01$, whilst keeping the relationship $Pe = 1/(9.27 Cn)$. We investigated turbulence transition in a range of pipe lengths, $L_{z,1} = \pi/2$, $L_{z,2} = \pi$, $L_{z,3} = 2\pi$ and $L_{z,4} = 4\pi$, which are denoted as $S_1$, $S_2$, $S_3$ and $S_4$, respectively. The DNS were initialized with PCAF, which was disturbed with a small perturbation exciting all Fourier modes. This perturbation consisted of a snapshot of a turbulent single-phase flow simulation at $Re=6000$, which was rescaled so that its total maximum energy was $<10^{-8}$.  

In the Cahn--Hilliard method, the interface thickness is not influenced by the grid resolution, but it is set uniquely by the Cahn number $Cn$, which is the ratio between the interface and bulk free energies. The grid resolution must then be selected to resolve the profile of $C$ across the interface properly ($\ge5$ points across the interface). Because the initial state is close to the laminar profile, the interface is initially approximately parallel to both axial and azimuthal directions. Therefore, a relatively small initial number of modes was used in both directions, with $n_r=96$, $2M=64$ and $2K=512$ for $S_4$, and a proportionally smaller number of axial modes for the shorter pipes. The grid resolution was successively increased as the simulation evolved and the interface became increasingly non-parallel to the wall parallel directions. This deformation produced an increase of the energy of the smallest resolved scales, which was used as the control parameter for the resolution. The energy was monitored at runtime and the resolution was increased whenever the energy of the smallest amplitude Fourier mode exceeded $10^{-10}$ times the energy of the highest amplitude mode. The time-step size was $\Delta t = 5\cdot 10^{-4}$ to ensure that the fast interface dynamics were properly captured \cite{Song2019}. 

\subsection{Weakly nonlinear regime}

\begin{figure}[!ht]
\begin{subfigure}[c]{0.45\textwidth}
\includegraphics[scale = 1]{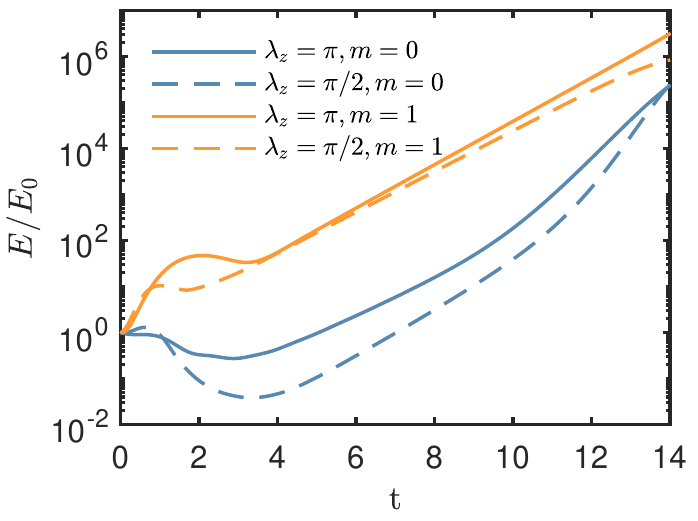}
\subcaption{ } \label{fig:DNS_mode_energy_S2}  
\end{subfigure}
\hspace{5mm}
\begin{subfigure}[c]{0.45\textwidth}
\includegraphics[scale = 1]{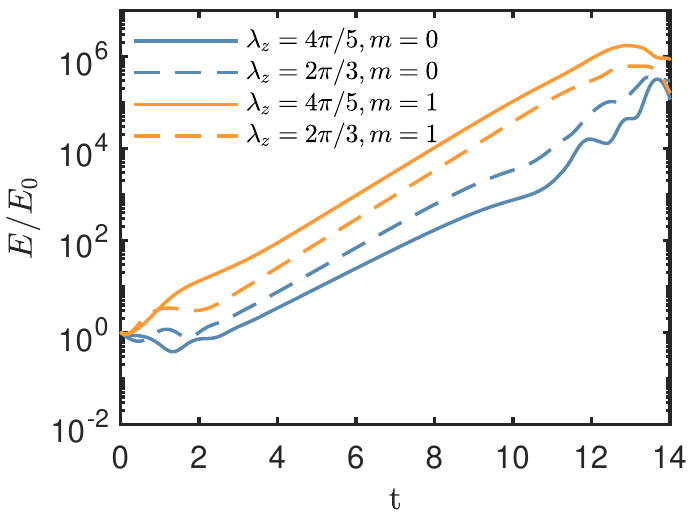}
\subcaption{ } \label{fig:DNS_mode_energy_S4}  
\end{subfigure}
\caption{Temporal evolution of most energetic modes in each of the DNS simulations. a) $S_2$ and b) $S_4$. Blue lines correspond to axisymmetric modes $m=0$ and yellow lines to $m=1$ non-axisymmetric modes. 
 } \label{fig:DNS_mode_energy} 
\end{figure}

The temporal evolution of the energy of the fastest growing modes is shown in figure~\ref{fig:DNS_mode_energy} for the runs $S_2$ and $S_4$. Initially, the non-axisymmetric modes show a significant non-modal behavior before they settle into the asymptotic state characterized by an exponential energy growth. Overall, the behavior is in agreement with the predictions of the linear stability analysis performed in section~\ref{sec:stability}. In all cases the non-axisymmetric $m=1$ modes with axial  wavelengths $\lambda_z\in[\pi/2,\pi]$ are the most unstable. In the run with the longest pipe ($S_4$, shown in  figure~\ref{fig:DNS_mode_energy_S4}), several modes with similar growth rates compete. As a consequence, the initial non-modal growth and the perturbation choice play both an important role in determining the mode with highest energy in the transition process. At $t\approx 12$, the growth of the perturbations begins to saturate and nonlinear effects begin to dominate; note that in run $S_2$ nonlinear effects kick in slightly later.

 \begin{figure}
\centering
\begin{subfigure}[c]{0.2\textwidth}
\includegraphics[width=0.8\textwidth]{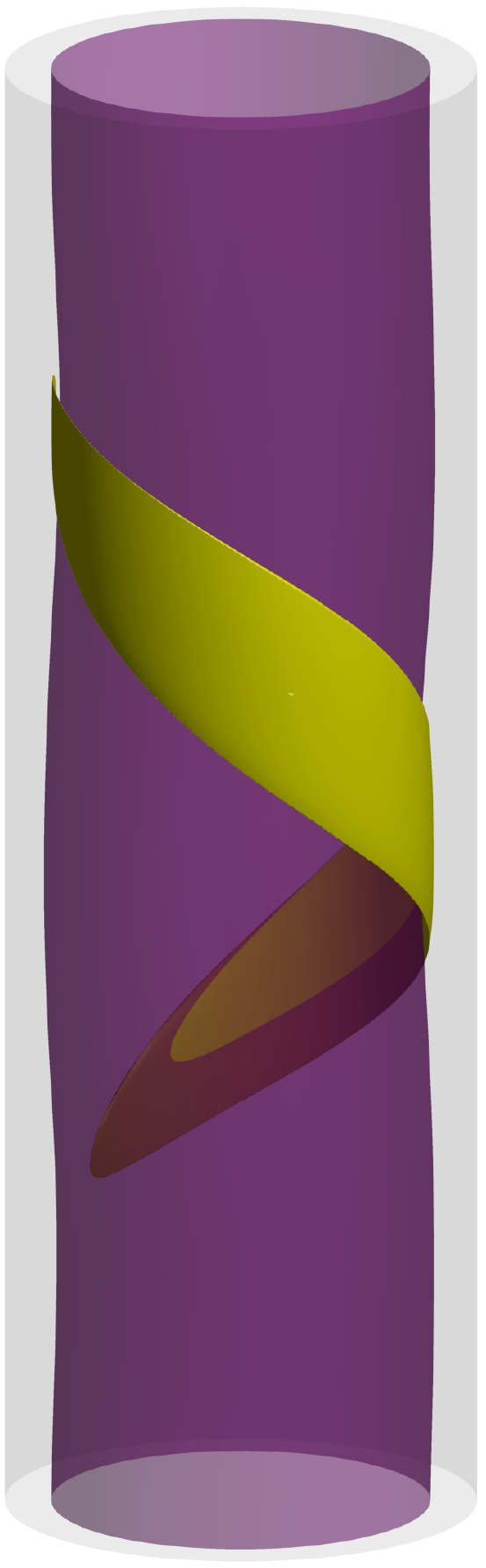}
\subcaption{t = 11.8} \label{fig:early_perturbation_3} 
\end{subfigure}  
\begin{subfigure}[c]{0.2\textwidth}
\includegraphics[width=0.8\textwidth]{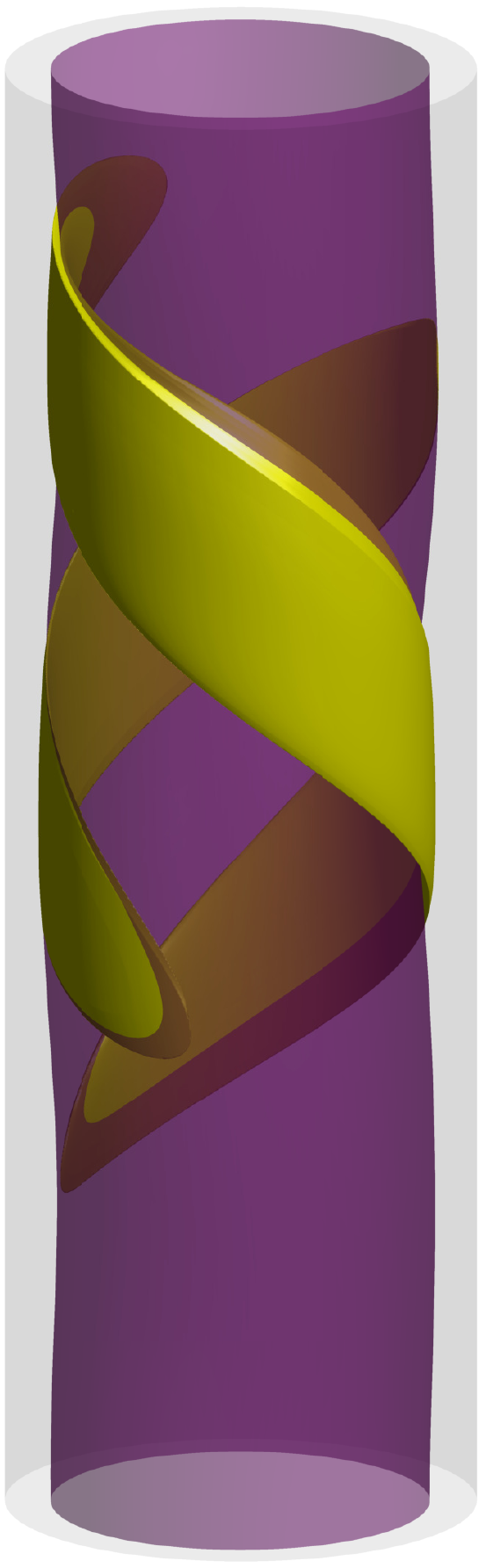}
\subcaption{t = 12.5} \label{fig:early_perturbation_4}  
\end{subfigure}
\begin{subfigure}[c]{0.2\textwidth}
\includegraphics[width=0.8\textwidth]{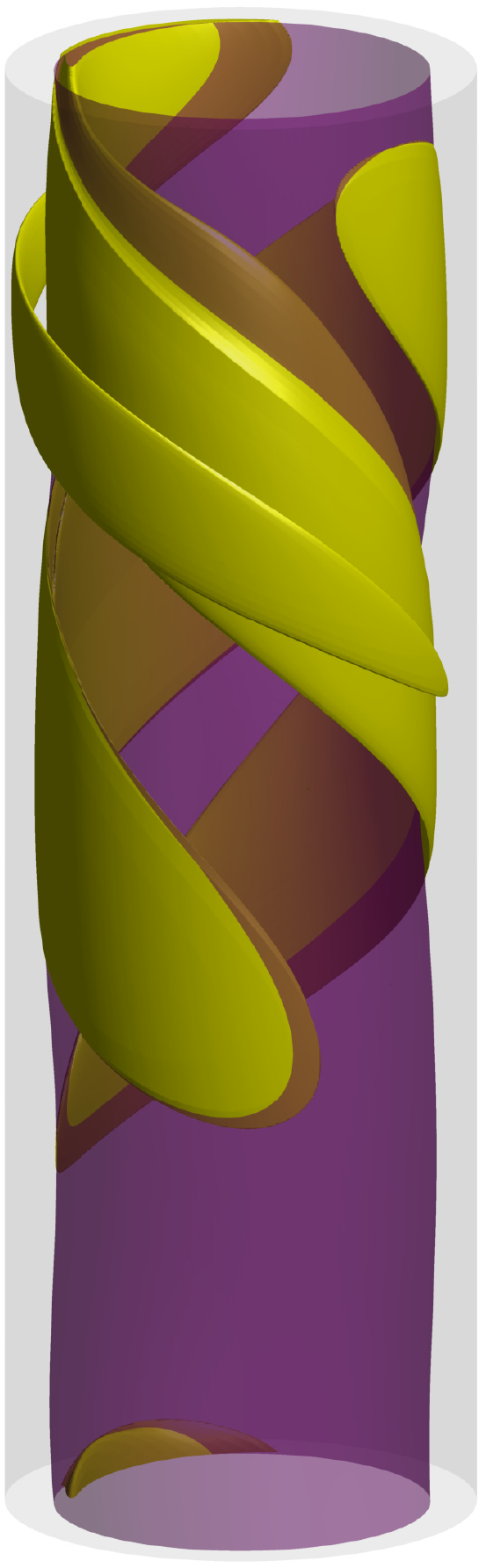}
\subcaption{t = 13.1} \label{fig:early_perturbation_5} 
\end{subfigure} 
\begin{subfigure}[c]{0.2\textwidth}
\includegraphics[width=0.8\textwidth]{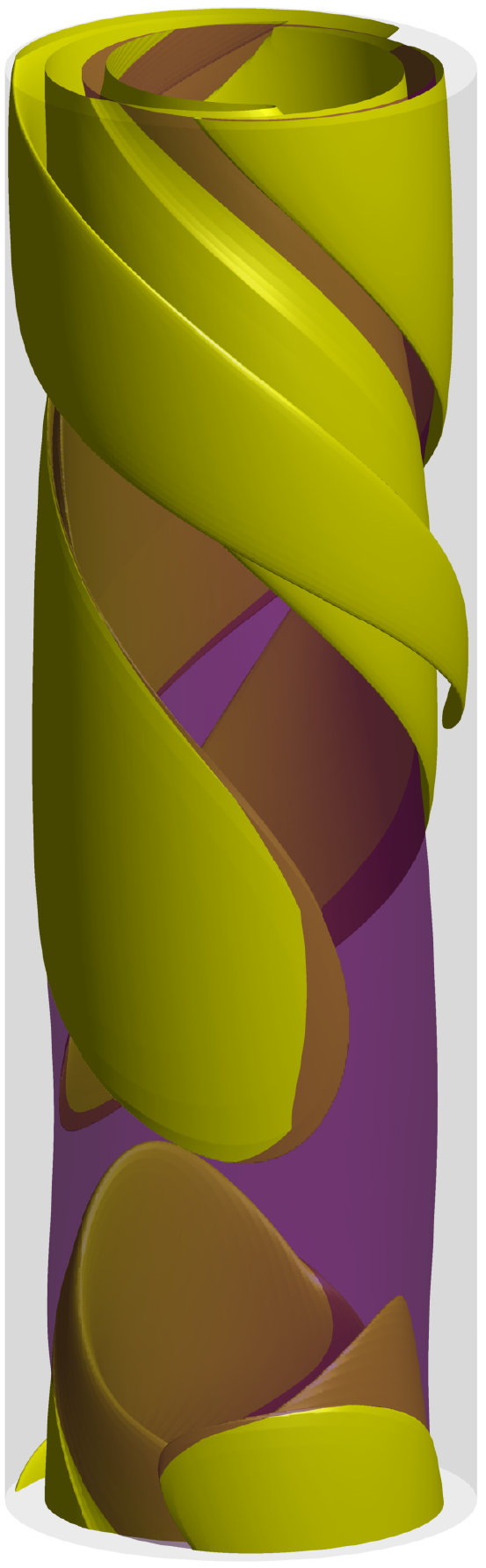}
\subcaption{t = 13.7} \label{fig:early_perturbation_6} 
\end{subfigure}  
\caption{Development of spiral structures on top PCAF of run $S_2$. Purple: Fluid interface. Yellow: iso-contour of the perturbation velocity, $\vec{u}_p$ in eq.~\eqref{eq:pert_vel}, with $|\vec{u}_p| = 0.195$. 
}   \label{fig:early_perturbation} 
\end{figure}

In figure~\ref{fig:early_perturbation}, we show isosurfaces of the perturbation velocity, 
\begin{align} \label{eq:pert_vel}
    \vec{u}_p(r,\theta, z) &= 
    	\begin{bmatrix}
			u_z(r,\theta, z) - u_{z,PCAF}(r) \\
          	u_\theta(r,\theta, z) \\
          	u_r(r,\theta, z)
         \end{bmatrix}.
\end{align}
and the interface during the early stages of the run $S_2$. The original cylindrical interface characteristic of the PCAF is deformed by non-axisymmetric traveling waves of increasing amplitude. The resulting structure is dominated by the spiral mode $(\lambda_z,m) = (\pi,1)$, but the contribution of the mode  $(\lambda_z,m) = (\pi/2,1)$ is also noticeable and leads to an imperfect spiral structure. The other runs show the development of similar spiral structures.

\subsection{Break-down of CAF configuration}

\begin{figure}[!ht]
\begin{subfigure}[c]{1\textwidth}
\includegraphics[scale=1]{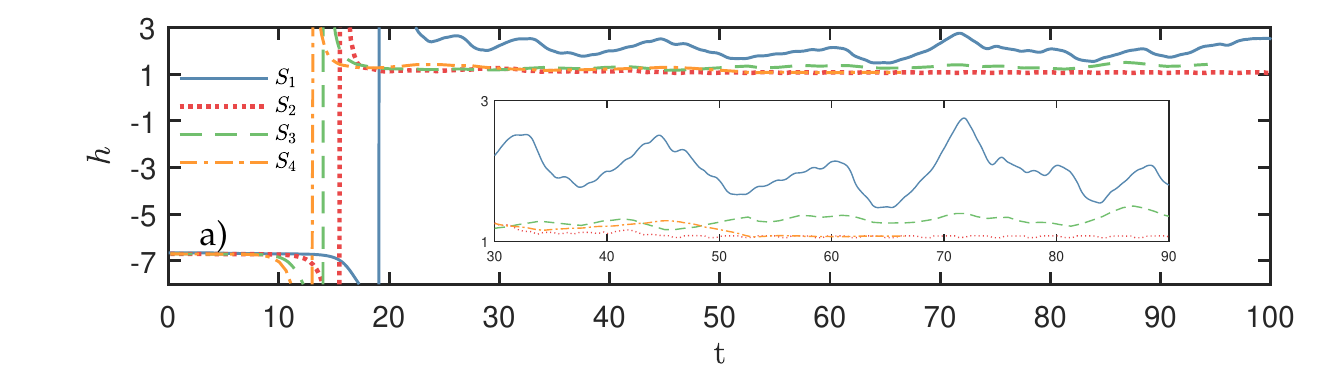}
\phantomsubcaption{\label{fig:hold-up}  }
\end{subfigure}
\\
\begin{subfigure}[c]{1\textwidth}
\includegraphics[scale=1]{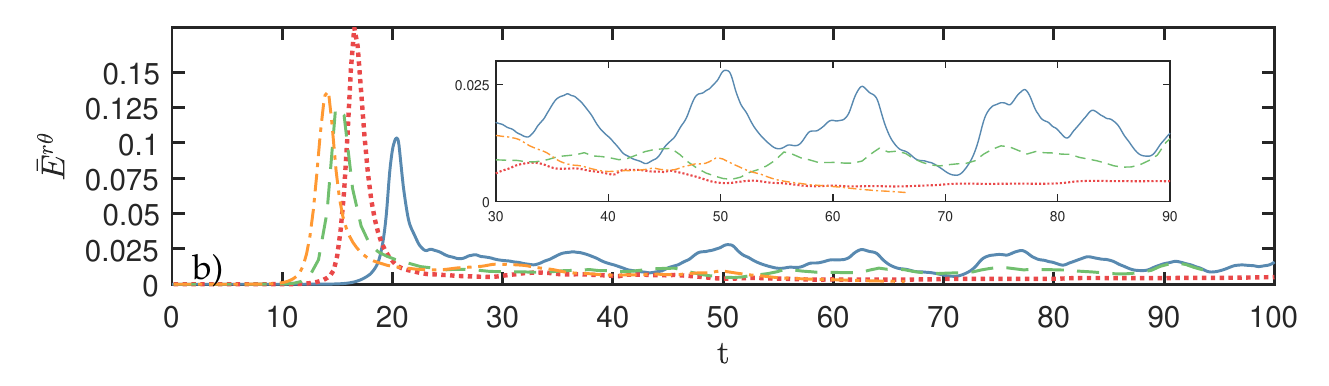}
\phantomsubcaption{ \label{fig:av_KE} }
\end{subfigure}
\\
\begin{subfigure}[c]{1\textwidth}
\includegraphics[scale=1]{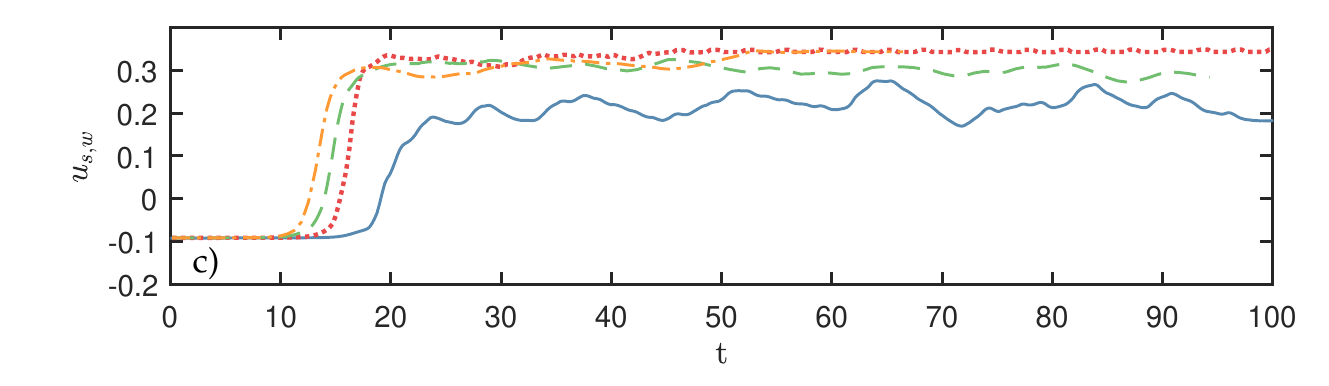}
\phantomsubcaption{\label{fig:av_sup_vel} } 
\end{subfigure} 
\\
\begin{subfigure}[c]{1\textwidth}
\includegraphics[scale=1]{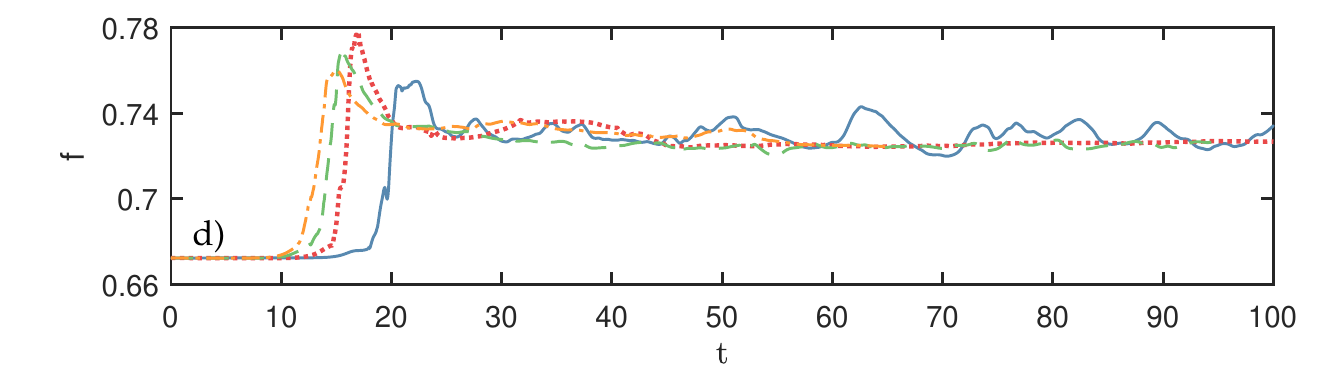}
\phantomsubcaption{\label{fig:pressure_gradient} } 
\end{subfigure}

\caption[]{\label{fig:time_evol} $S_1 (L_z = \pi/2)$, $S_2 (L_z = \pi)$, $S_3$ $(L_z = 2\pi)$, $S_4$ $(L_z = 4\pi$) a) Hold up ratio ($h$). b) Average kinetic energy of the mixture ($\bar{E}^{r\theta}$). c) Water superficial velocity ($u_{s,w}$). d) Driving pressure gradient, ($f$). Inserts in a) and b) depict a close-up of the saturated region. }
        
    \end{figure}

The evolution of the hold-up ratio, eq.~(\ref{eq:hold-up-ratio}), during all simulations is shown in figure~\ref{fig:hold-up}. All cases follow the same basic pattern. During the early stages the perturbation is still small and,  while it grows exponentially, its influence on the basic PCAF profile is negligible and the hold-up ratio remains basically constant (linear regime). Once the perturbation grows past a certain threshold, the flow transitions violently from the original PCAF into a new saturated state. During the transition the resolution reaches a maximum, namely $n_r=96$, $2M=768$ and $2K=3072$ for $S_4$ and similar values, proportionally reduced in the axial direction for the shorter pipes. In this phase, the water flow rate changes sign from the negative value of the PCAF (due to the downwards pull of gravity) into a positive value, as a result of the enhanced momentum transfer from the oil phase, causing a brief divergence of the hold-up ratio. The starting point of the transition phase depends on the pipe length, with longer pipes showing earlier transitions, because the increased length means they can accommodate modes closer to the peak in the growth rate profile (see figure~\ref{fig:laminar_gr}).

 \begin{figure} 
 \centering
\begin{subfigure}[c]{0.2\textwidth}
\includegraphics[width=0.8\textwidth]{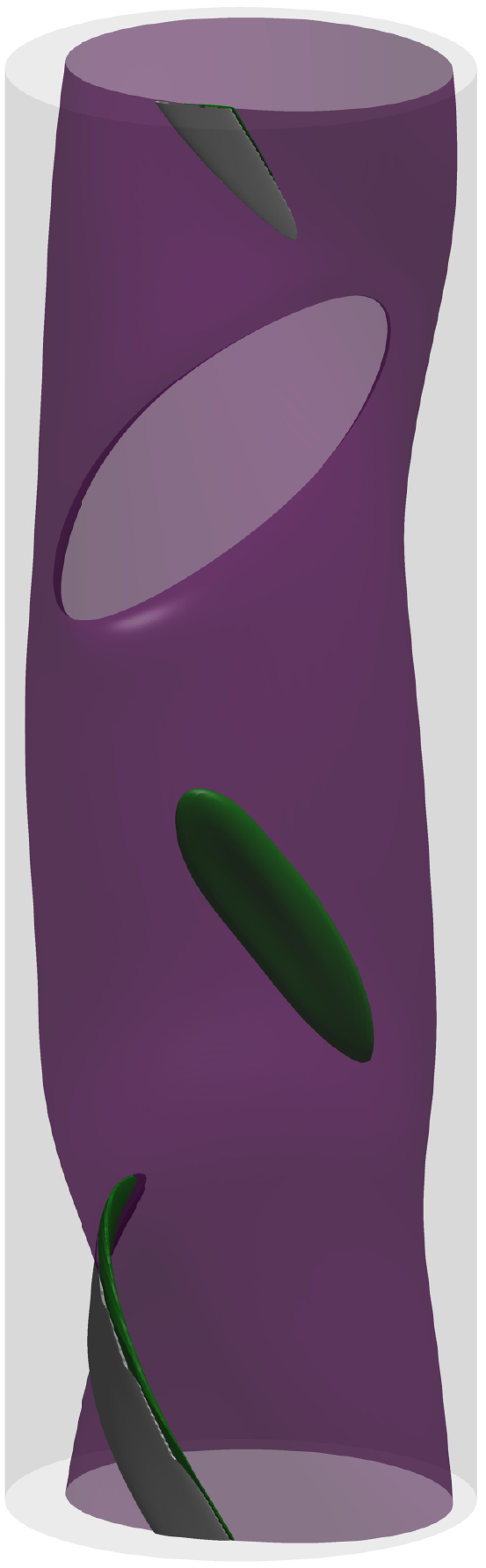}
\subcaption{t = 14.7} \label{fig:CAF_transition_1} 
\end{subfigure} 
\begin{subfigure}[c]{0.2\textwidth}
\includegraphics[width=0.8\textwidth]{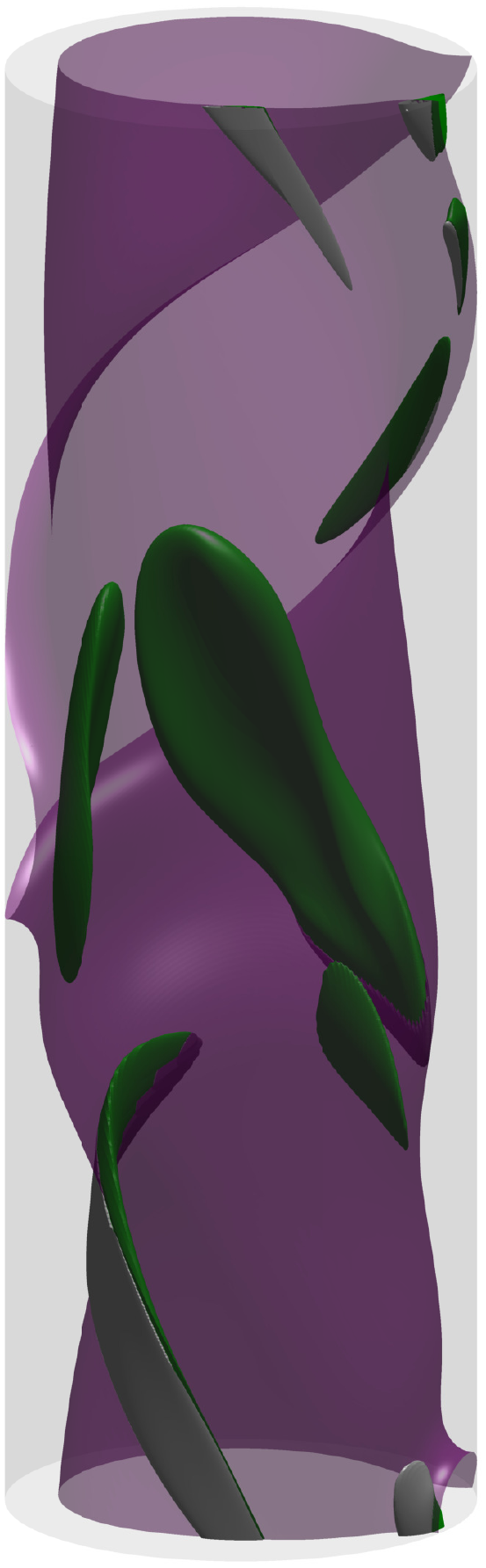}
\subcaption{t = 15.0} \label{fig:CAF_transition_2} 
\end{subfigure} 
\begin{subfigure}[c]{0.2\textwidth}
\includegraphics[width=0.8\textwidth]{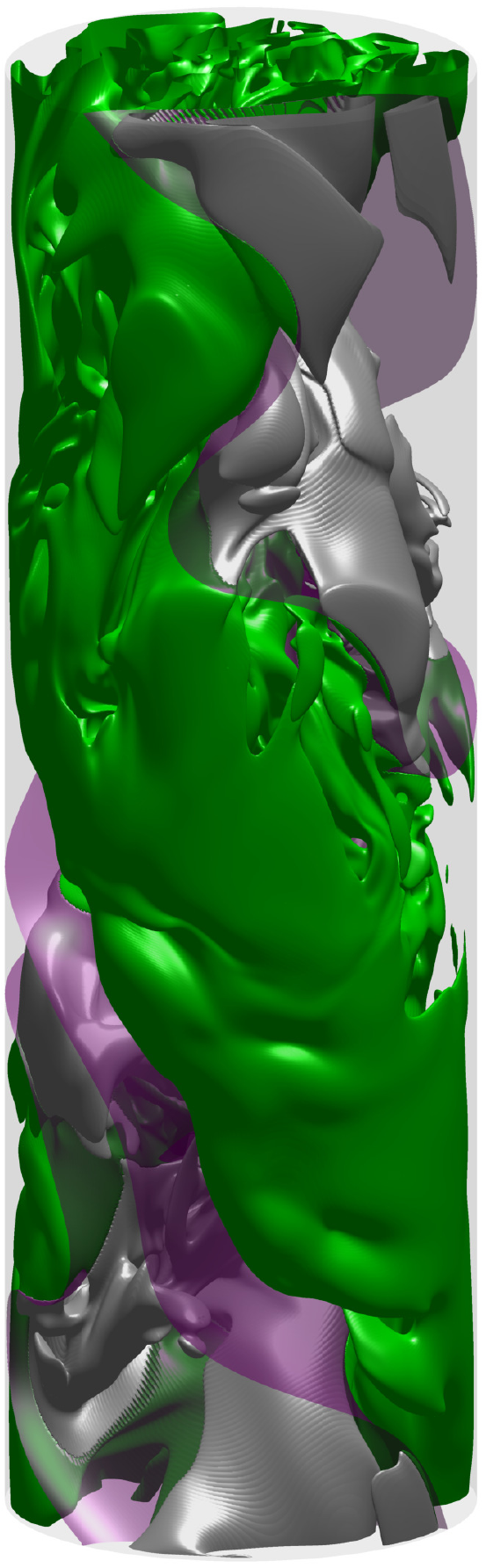}
\subcaption{t = 16.8} \label{fig:CAF_transition_3} 
\end{subfigure} 
\begin{subfigure}[c]{0.2\textwidth}
\includegraphics[width=0.8\textwidth]{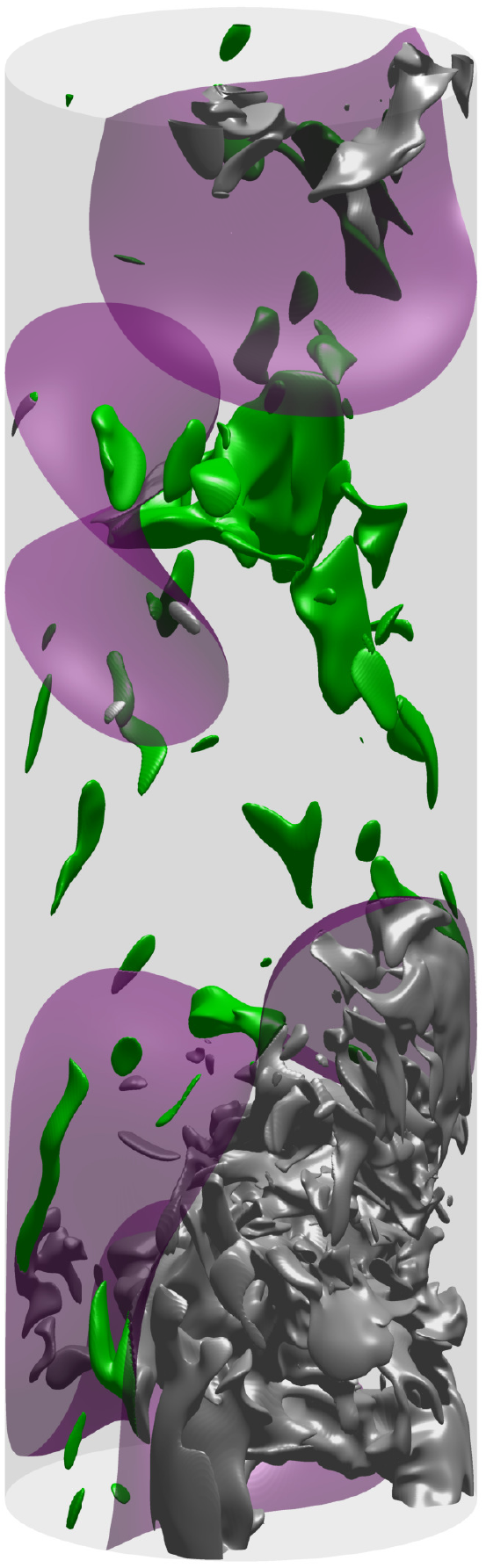}
\subcaption{t = 18.7} \label{fig:CAF_transition_4} 
\end{subfigure} 

\caption{Break-up of PCAF into turbulent slugs in run $S_2$. Purple: Fluid interface. Green/Gray: isosurfaces of $E^{r\theta} = 0.11$ (oil/water).} 
\label{fig:CAF_transition} 
\end{figure} 
To characterize the chaotic motion occurring in this stage, and later stages, a measure of the turbulent kinetic energy is desired. In single-phase flow, the turbulent kinetic energy is obtained in two steps. First time and space (in the axial and azimuthal direction for pipe flow) averages are taken to calculate the mean profile. This profile is then subtracted from the instantaneous velocity and the kinetic energy is calculated from the fluctuation. Due to the nature of multiphase flows, the calculation of the mean velocity field of an arbitrary multiphase flow is not useful, since the topological changes and the continuous displacement of the interface make the usual averaging procedure meaningless, both in space and time. Not only does the distance to the interface of any particular point drastically change with time, but the very properties of the flow at that point may change due to the movement of the interface. Therefore, under the assumption that the mean azimuthal and radial velocities are zero, the kinetic energy of the cross-sectional (in-plane) velocity 
 \begin{equation} \label{eq:2D_KE}
  E^{r\theta} = 0.5 \rho(C)(u^2_r + u^2_\theta).
 \end{equation}
 was used as a measure of the turbulence intensity of the oil and water phases. The volume-averaged kinetic energy was calculated as
 \begin{equation} \label{eq:av_KE}
  \bar{E}^{r\theta} = \frac{\int_{V} E^{r\theta} dV}{V}.
 \end{equation}
The break-down process in run $S_2$ is illustrated in figure~\ref{fig:CAF_transition}, where the interface and isosurfaces of $E^{r\theta}$ are displayed. The spiral structure developed during the weakly linear regime becomes so pronounced that the oil phase touches the wall and starts to slow down, causing a sharp increase in the turbulent kinetic energy (figures \ref{fig:CAF_transition_1}-\ref{fig:CAF_transition_3}). After a short time, the initial CAF structure splits into several smaller water entities surrounded by oil, and the intensity of the turbulence kinetic energy is greatly reduced (figure \ref{fig:CAF_transition_4}).

\subsection{Fully saturated state}

 \begin{figure} 
 \centering
\begin{subfigure}[c]{0.2\textwidth}
\includegraphics[width=0.8\textwidth]{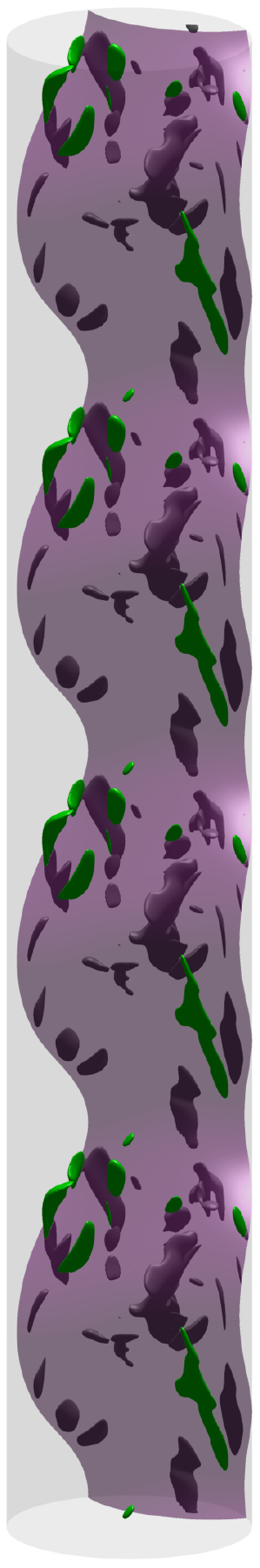}
\subcaption{$S_1$}  \label{fig:saturated_comparison_1}
\end{subfigure} 
\begin{subfigure}[c]{0.2\textwidth}
\includegraphics[width=0.8\textwidth]{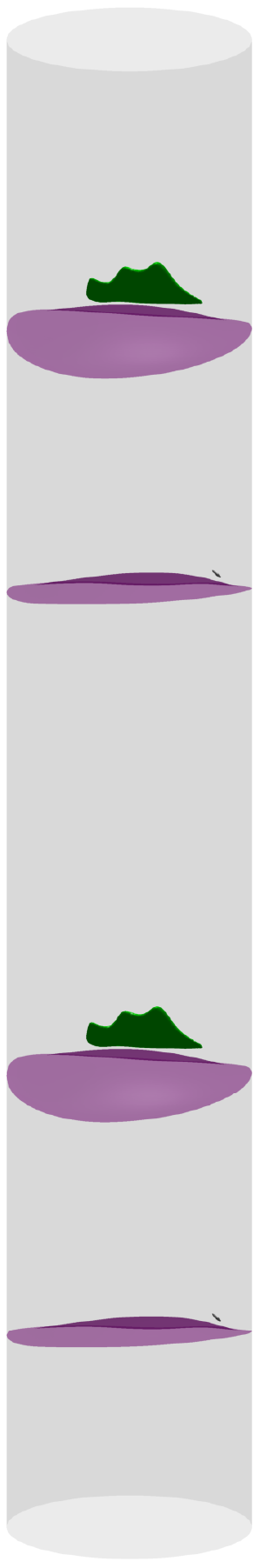}
\subcaption{$S_2$}  \label{fig:saturated_comparison_2}
\end{subfigure} 
\begin{subfigure}[c]{0.2\textwidth}
\includegraphics[width=0.8\textwidth]{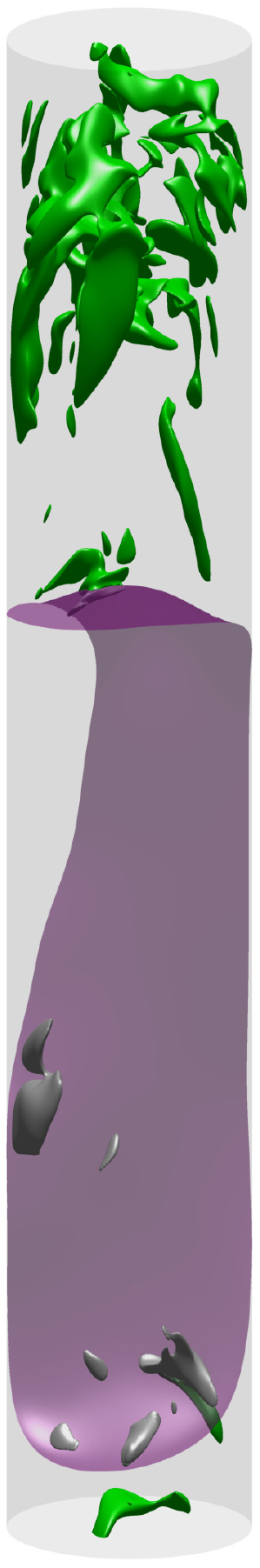}
\subcaption{$S_3$} \label{fig:saturated_comparison_3}
\end{subfigure} 
\begin{subfigure}[c]{0.2\textwidth}
\includegraphics[width=0.8\textwidth]{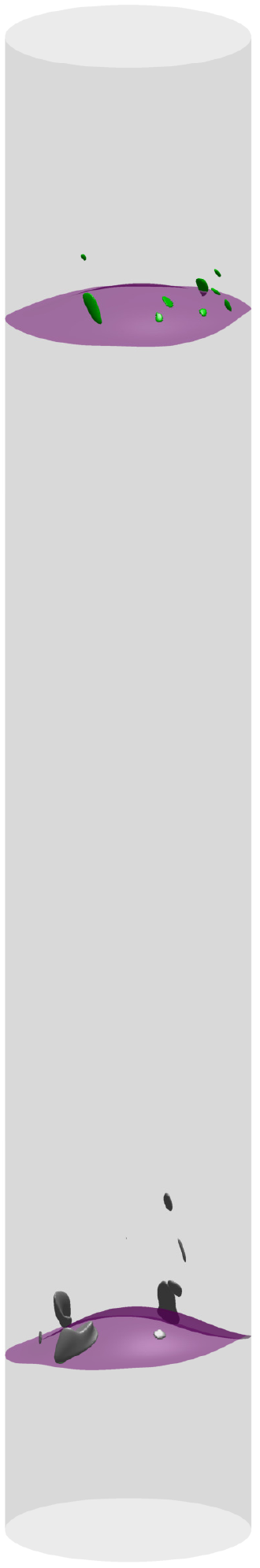}
\subcaption{$S_4$} \label{fig:saturated_comparison_4}
\end{subfigure} 

\caption{Isosurfaces of the saturated state for simulations $S_1$ - $S_4$. Purple: Fluid interface. Green/Gray: $E^{r\theta} = 0.06$ (Oil/Water). The flow fields of $S_1$ and $S_2$ are replicated four times and twice, respectively. In the view of $S_4$ only half the pipe length is shown (zoomed around the water slug structure). } \label{fig:saturated_comparison}
\end{figure}

After the transition to turbulence, the system reached a final saturated state, which depended on the pipe length, as shown in figure~\ref{fig:saturated_comparison}. The turbulence intensity settled down to lower levels and accordingly, the number of Fourier modes was reduced, to $n_r=96$, $2M=384$, $2K=1536$ in the case $S_4$ and proportional values for the shorter pipes. All runs reached a statistically steady states, which are described in what follows. 

The shortest pipe, $S_1$ presents a stratified wavy state with both phases flowing next to each other, as shown in figure~\ref{fig:saturated_comparison}(a). This state has been reported for horizontal pipes in previous works \cite{Shi2017}. On the other hand, the longer pipes exhibit two distinctly different end states which are displayed in figures~\ref{fig:saturated_comparison}(b)--(d). Both $S_2$ and $S_4$ converged into a configuration, where the water and the oil are distributed axially into slugs and the water occupies the whole cross-section of the pipe. By contrast, $S_3$ converged into a similar structure with a single water drop, with the difference that in this case the water does not fill the complete pipe cross-section, allowing for the concurrent flow of water and kerosene. 
  
The inset in figure~\ref{fig:av_KE} shows the evolution of the averaged kinetic energy of the mixture after the break-down of the CAF. The stratified wavy state found in $S_1$ exhibits large-amplitude oscillations, which consist of alternate phases in which turbulence is intense and the interface flattens, and phases in which the interface becomes wavier and the turbulence intensity decreases. In $S_3$, the water drop flowing alongside the kerosene matrix, presents a much smaller value of the kinetic energy and weaker oscillations. Finally, the slug flow configuration found in $S_2$ and $S_4$ quickly settles into a relatively ordered regime with a reduced value of kinetic energy. Similar patterns of large oscillations in the stratified flow of $S_1$, weaker ones in the drop configuration of $S_3$ and settling to constant values in the slug flow of $S_2$ and $S_4$ can be seen in the other flow properties, such as the water superficial velocity, $u_{s,w}=Q_w/A$  and the driving pressure gradient $f$, as shown in figures~\ref{fig:av_sup_vel} and \ref{fig:pressure_gradient}. In the case of $u_{s,w}$ we can see the change of direction in the mean water flow rate from flowing downwards (negative $u_{s,w}$) in CAF  to flowing upwards (positive $u_{s,w}$) in all cases once the flow becomes saturated. Due to the constraint of constant volumetric flow rate, this entails a reduction in the oil flow rate, $u_{s,o} = 1 - u_{s,w}$. 

It is worth noting that the main contribution of the driving pressure gradient counteracts the gravity force pulling the fluid downwards, averaging $f_g = -0.694$ over the whole pipe, and only a fraction of it can be attributed to friction losses at the wall. However, because $f_g$ is constant in time, any change in the pressure gradient reflects changes in the wall friction. The PCAF presents a negative value of friction (note that initially $f < 0.694$), due to the negative axial velocity of the water phase close to the wall. When the transition starts, there is a sudden increase in the pressure drop (and in the corresponding friction force), caused by the change of the water velocity profile close to the wall and the migration to the wall region of fast moving oil, with a higher viscosity. Once the transition is completed, it settles into a smaller value, with little difference between the slug and drop regimes, meaning that both configurations suffer from comparable wall friction loses. 
 
\subsection{Drops and slugs}
 
\begin{figure}
\captionsetup[subfigure]{justification=centering}
\begin{subfigure}[c]{0.32\textwidth}
\includegraphics[width=0.99\textwidth]{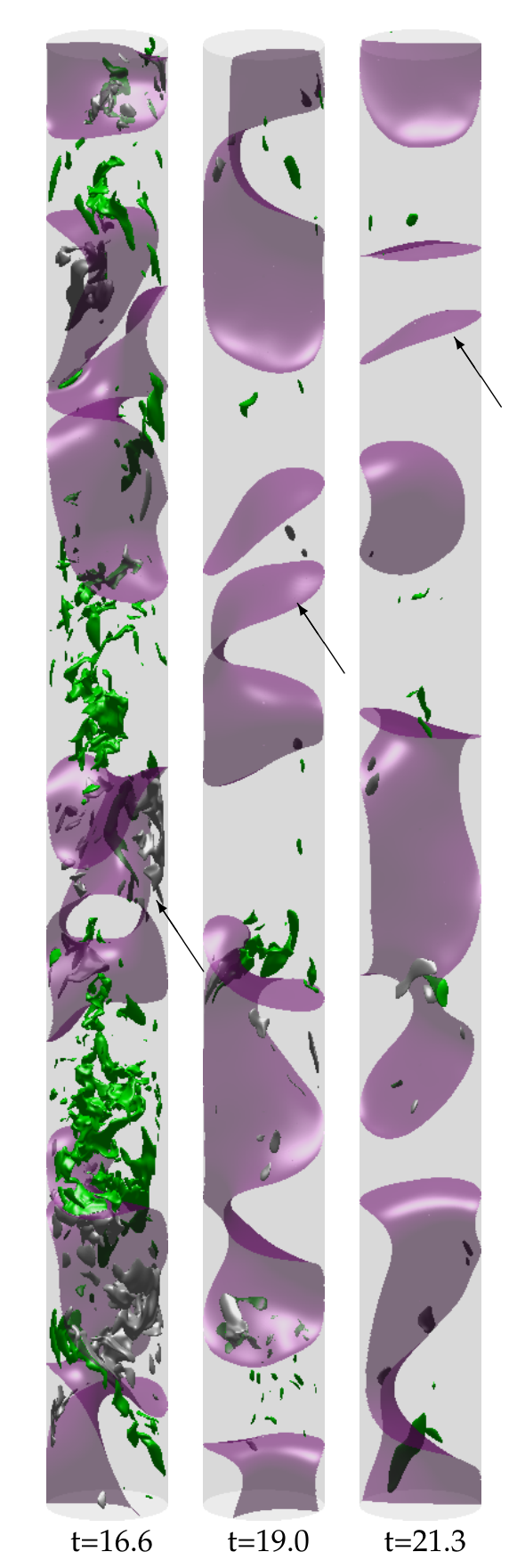}
\subcaption{Formation of initial slug} \label{fig:slug_formation_S4_1} 
\end{subfigure} 
\begin{subfigure}[c]{0.32\textwidth}
\includegraphics[width=0.99\textwidth]{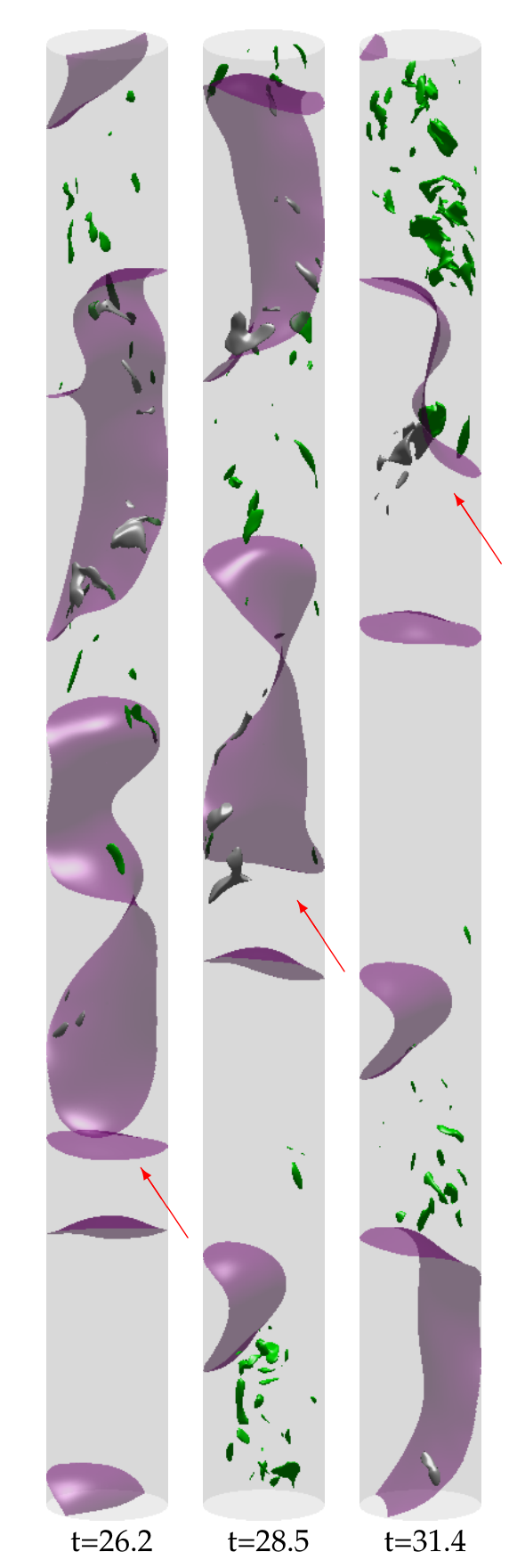}
\subcaption{First interaction drop/slug} \label{fig:slug_formation_S4_2} 
\end{subfigure} 
\begin{subfigure}[c]{0.32\textwidth}
\includegraphics[width=0.99\textwidth]{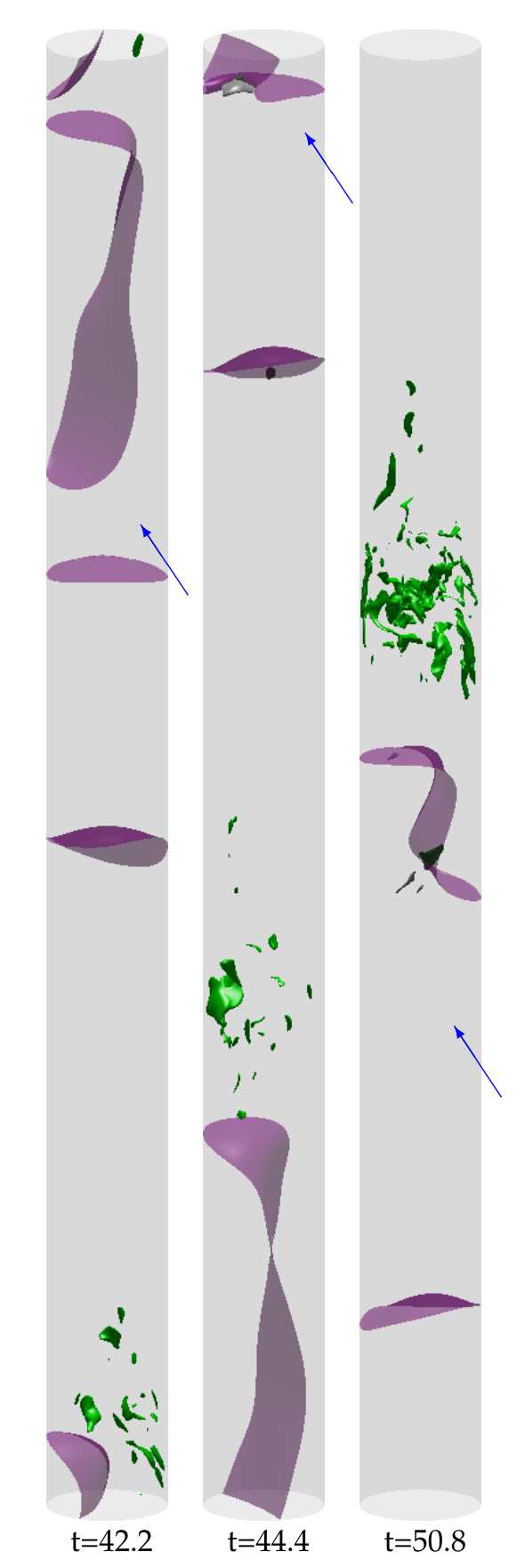}
\subcaption{Second interaction drop/slug} \label{fig:slug_formation_S4_3} 
\end{subfigure} 
\caption{Process of slug formation in $S_4$. a) Formation of the initial slug after CAF break down. b)First occurrence of drop/slug interaction. c) Second occurrence of drop/slug interaction. Purple: Fluid interface. Green/Gray: $E^{r\theta} = 0.11$ (Oil/Water). Arrows in each group (a,b,c) point towards the same fluid region in the moving reference frame of the upflow periodic pipe.
 \label{fig:slug_formation_S4}}
\end{figure}

The formation process for the final slug structure in run $S_4$ is shown in figure~\ref{fig:slug_formation_S4};  $S_2$ follows a similar structure and is not shown here. After the break down of the CAF, several smaller structures are formed, including a small body of water that covers the whole pipe cross section. This is the initial slug and is marked with a black arrow in figure~\ref{fig:slug_formation_S4_1}. Surface tension forces keep the slug-kerosene interface stable, and therefore, due to mass conservation it is forced to travel along the pipe at the same speed as the kerosene immediately before and after it. In contrast, the water drops that only partially fill the pipe cross-section travel slower than the surrounding kerosene, because water is less buoyant.  As a consequence, the water slug catches up with the preceding water drops and absorbs them. This process is highlighted with red and blue arrows in figures \ref{fig:slug_formation_S4_2}  and \ref{fig:slug_formation_S4_3} and can also be interpreted as the drainage of the oil buffer between the slug and the drop. The resulting state thus consists of a water slug and a kerosene slug.

\begin{figure}[!ht]
        \centering
            \centering 
            {{\small }}   
            \includegraphics[scale=1]{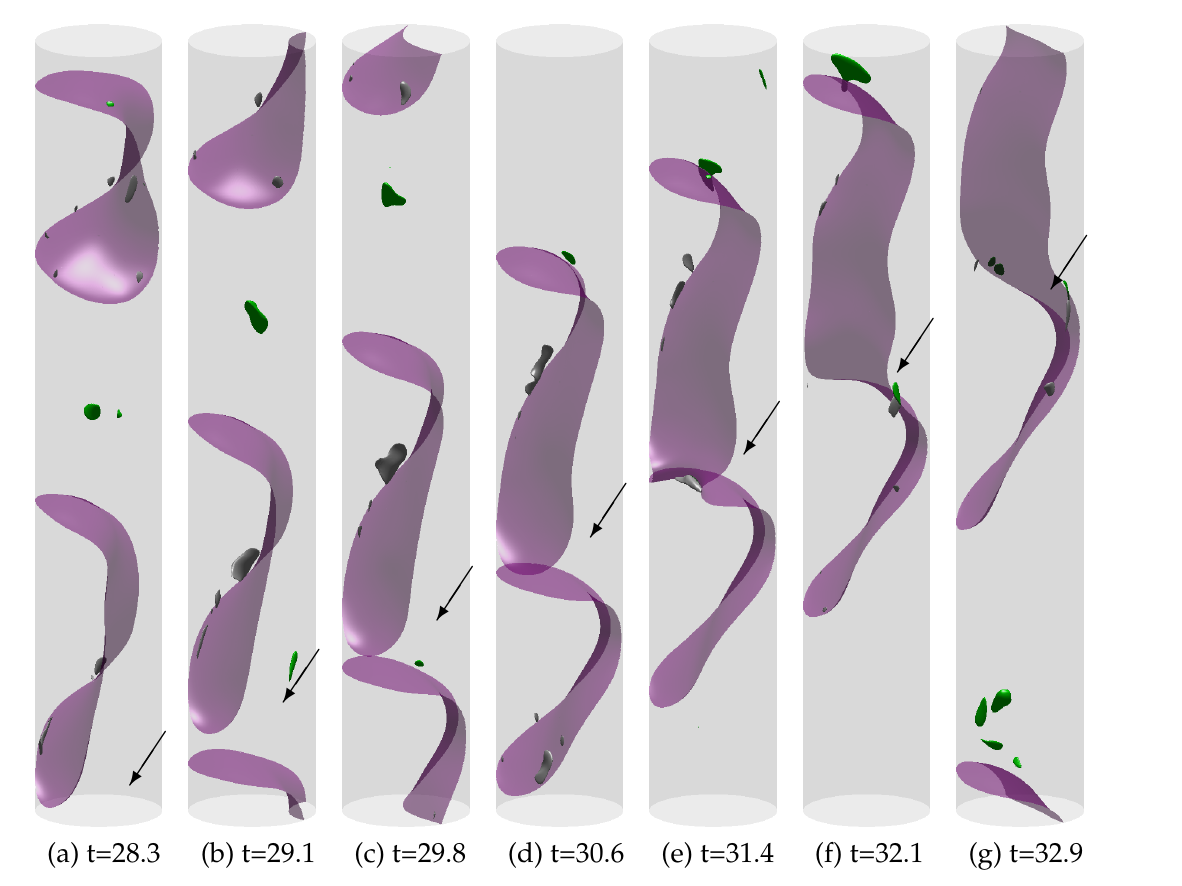}
        \caption[]{\label{fig:drop_formation_S3} 
Process of coalescence of individual water drops in $S_3$ after the initial CAF break down is complete. Purple: Fluid interface. Green/Gray: $E^{r\theta} = 0.11$ (Oil/Water). Black arrows point towards the same fluid region in the moving reference frame of the upflow periodic pipe.
}    
\end{figure}

Figure \ref{fig:drop_formation_S3} shows the formation process of the stable single water drop that constitutes the saturated state in $S_3$. In this case the break down of the CAF did not give rise to any slug, but to several drops. These drops flowed upwards dragged by the surrounding kerosene, with different speeds depending on their shape and size. In a similar manner as in the slug formation, these drops sooner or later interact with each other, forming larger drops until a single stable drop remains. However, in contrast to the previous case, here the water is not able to cover the whole pipe cross-section and kerosene flows alongside the water. This single drop configuration remains stable throughout the reminder of the run and is a less ordered state than the slug: the interface has a larger surface area and is therefore more easily deformable while at the same time being able to rotate in the cross-sectional plane. The effect of these differences can be appreciated in figure \ref{fig:time_evol}, where the drop configuration has a consistently larger value of the hold-up ratio $h$ and $\bar{E}^{r\theta}$ and a more chaotic behavior than the slug flow. 

\subsection{Influence of initial conditions}

We have shown that three distinctive flow configurations may be achieved by starting from disturbed PCAF and varying the pipe length. In this section, we investigate the effect of initial conditions systematically. For this purpose, we took the saturated states obtained so far, and exploiting the periodic boundary conditions, we appended two identical copies of the structure in a pipe with double axial extent. Then the flow was left to evolve freely.

The robustness of the wavy stratified flow found in $S_1$ was tested by running $S_{2b}$ in a pipe of length $L=\pi$, as employed in run $S_2$. The initial condition consisted of a snapshot of the stratified flow of $S_1$ at $t=31$, with the complete flow structure replicated twice in the axial direction. After a long relaxation time, of more than $70$ advective time units, the wavy interface broke down and the system converged into a single water drop, analogous to the one shown in figure \ref{fig:saturated_comparison_3}, but with a smaller water drop size due to the shorter pipe length. This confirms that the stratified flow in $S_1$ is an artifact resulting from the short pipe length.

\begin{figure}
\begin{center}

\captionsetup[subfigure]{justification=centering}
\begin{subfigure}[c]{0.40\textwidth}
\includegraphics[width=0.99\textwidth]{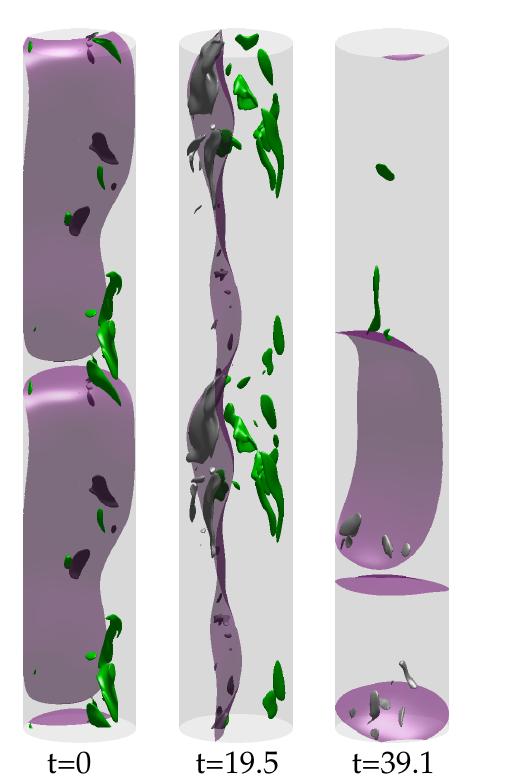}
\subcaption{Formation of slug from drop interactions} \label{fig:drop_stability_S3_b} 
\end{subfigure} 
\begin{subfigure}[c]{0.40\textwidth}
\includegraphics[width=0.99\textwidth]{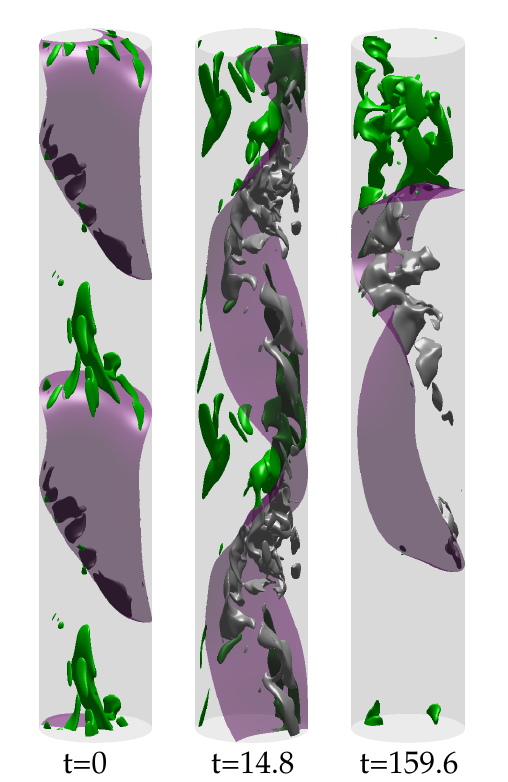}
\subcaption{Formation of drop form drop interaction} \label{fig:drop_stability_S3_c}

\end{subfigure} 

\caption{Two examples of drop interaction process in a pipe of $L = 2\pi$ initialized with the twice-repeated flow structure of $S_{2b}$. Green/Gray: $E^{r\theta} = 0.06$ (Oil/Water). a) Snapshots of the run $S_{3b}$, giving rise to the slug configuration. b) Snapshots of the run $S_{3c}$, resulting in a larger water drop. The time of the snapshots is referenced to the start of its own run.
 \label{fig:drop_stability_S3}}
 \end{center}
\end{figure}

To probe the stability of the drop configuration, two additional runs were performed employing snapshots of $S_{2b}$ at different times as initial conditions. Note that duplicating the drop pattern twice in axial direction resulted in a flow state with two drops, whereas in run $S_3$ (started from disturbed CAF) only one drop was present at the end. In run $S_{3b}$, the initial condition was a snapshot $30$ advective time units after the start of $S_{2b}$, when the water drop was already present, but still relaxing to equilibrium. Run $S_{3c}$ was initialized with the flow conditions at the end of the run, when it was in equilibrium. Both cases followed a similar pattern: the two initial drops interacted with each other to form a temporal continuous interface, similar to the stratified flow found in $S_1$. This structure is not stable and breaks down into drops again. In the case of $S_{3b}$ one of the water drops manages to occupy the complete pipe cross-section, becoming a slug, which proceeds to absorb the remaining drops in the same way described in figure \ref{fig:slug_formation_S4}. In the case of $S_{3c}$, the single water drop arising from the collapse of the continuous interface remains as a drop for the complete run. This process can be visualized in figure \ref{fig:drop_stability_S3}. Figures \ref{fig:drop_stability_S3_b} and \ref{fig:drop_stability_S3_c}, corresponding to runs $S_{3b}$ and $S_{3c}$ respectively, show the initial conditions, the formation of the stratified structure and, in the former case the moment of the absorption of the last drop by the slug structure and in the later, the resulting larger water drop. 

The stability of the slug configuration was checked by using as initial condition the stable slug obtained from $S_{3b}$, again doubling the complete flow pattern in the axial direction. The resulting simulation is denoted $S_{4b}$. In this case, the pair of consecutive slugs was completely stable and no significant change in the flow structure or the flow properties was observed after more than $30$ advective times units. 

With this results in mind, it can be deduced that slug flow is the most likely configuration to be found under the flow conditions investigated in this paper. In experimental pipes, which are substantially longer than the ones used in our direct numerical simulations, we can expect that any arbitrary initial flow will eventually evolve into a mixture of slugs and drops. However, the drop configuration is only stable in isolation, with no other structure to interact with, which is not a very likely situation under real laboratory conditions. More likely, the slugs in the mixture will inexorably catch up with and absorb the neighboring drops, increasing in length. On the other hand, two neighboring slugs are not able to interact, because mass conservation forces them to travel at the same mean speed.

\subsection{Analysis of slug flow} \label{sec:analysis_slug}

The slug flow described in the previous sections presents a relatively stationary interface in the frame of reference moving with the slug, with only relatively small deformations of the interface separating both phases, caused by the interactions between the inertial forces of the chaotic flow and surface-tension forces. This simple structure, allows for the calculation of an approximate mean profile for each phase,
 \begin{equation} \label{eq:mean_vz}
\bar{u}_{z,o/w} (r,t) = \frac{\int_{A^{z\theta}_{o/w}} u_z(r,\theta,z,t) dA^{z\theta}}{A^{z\theta}_{o/w}(r,t)},
 \end{equation}
where $A^{z\theta}_{o/w}(r,t)$ is the area in the ($z,\theta$)-cylinder occupied by each phase at the radial position, $r$. The calculated radial profile is then averaged in time. This process is a very simple approximation, where we neglect the deformation of the interface and the relative position of each particular point with respect to the interfaces. For a more accurate representation, a two-dimensional velocity map depending on both the axial and radial position would be desired. However, that would require very long time series for a good statistical convergence of the mean velocities and the simple one-dimensional representation is enough to get some insight of the characteristics of the flow.

\begin{figure}[!ht]
\begin{subfigure}[c]{0.49\textwidth}
\includegraphics[scale=1]{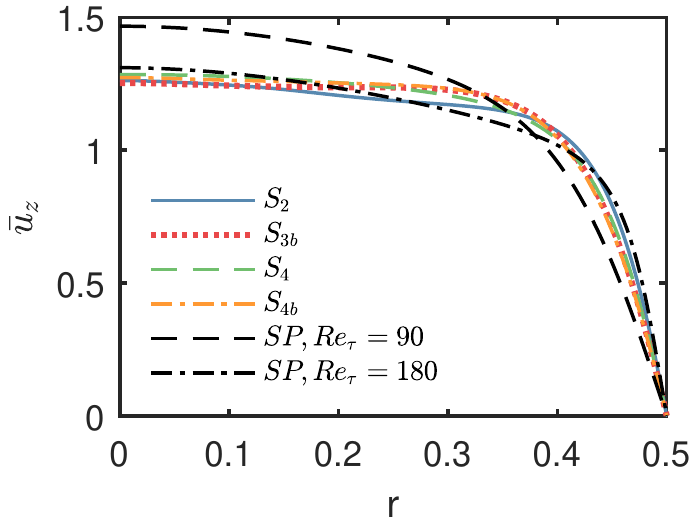}
\subcaption{ } \label{fig:DNS_slug_mean_flow_oil} 
\end{subfigure}
\begin{subfigure}[c]{0.49\textwidth}
\includegraphics[scale=1]{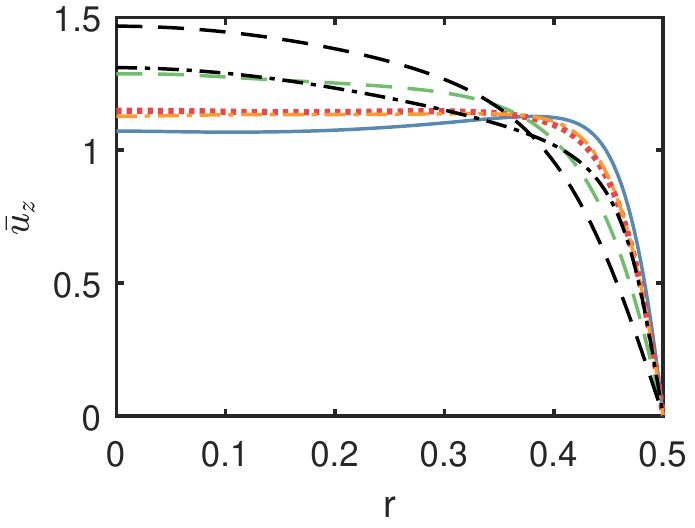}
\subcaption{} \label{fig:DNS_slug_mean_flow_wat}
\end{subfigure}
\begin{subfigure}[c]{0.49\textwidth}
\includegraphics[scale=1]{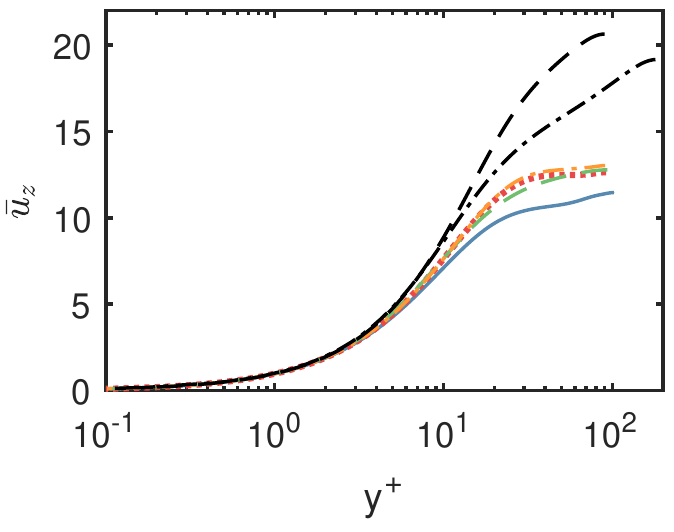}
\subcaption{ } \label{fig:DNS_slug_mean_flow_oil_yplus} 
\end{subfigure}
\hspace{0.21cm}
\begin{subfigure}[c]{0.49\textwidth}
\includegraphics[scale=1]{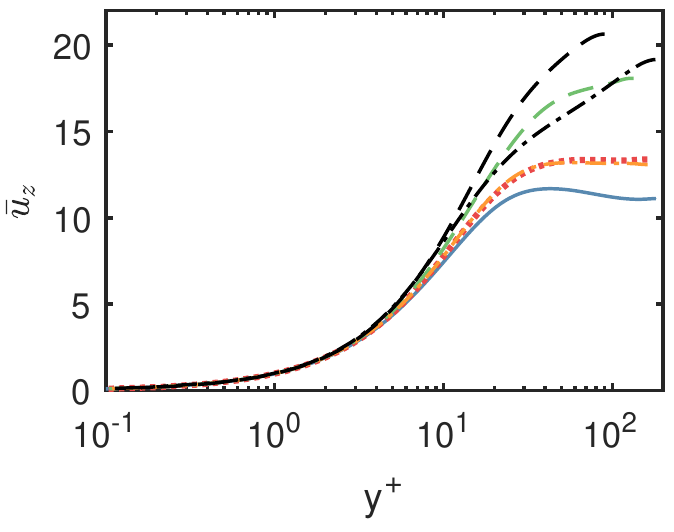}
\subcaption{} \label{fig:DNS_slug_mean_flow_wat_yplus}
\end{subfigure}

\caption{Radial mean profiles in the slug flow regime. a) Oil phase in bulk units b) water phase in bulk units. c) Oil phase in wall units. d) Water phase in wall units. Black dashed (dash-dot) line correspond to single-phase mean profiles at $Re_\tau=90$(180) taken from \cite{Feldmann2018}. $Re_\tau$ values for slug flow are $Re_{\tau,w} (S_2,S_{3b},S_4, S_{4b}) = [181, \, 161,\, 161, \, 131]$ for the water phase and $Re_{\tau,o} (S_2,S_{3b},S_4, S_{4b}) = [102,\, 91,\, 90,\, 93]$ for the oil phase.} \label{fig:DNS_slug_mean_flow} 
\end{figure}

The time-averaged mean profiles of axial velocity are plotted in figure \ref{fig:DNS_slug_mean_flow} for the four cases ($S_2$, $S_{3b}$, $S_4$, $S_{4b}$) that converged into slug flow. Alongside them are plotted the mean profiles of two single phase turbulent pipe flows, taken from \cite{Feldmann2018}, with  $Re_\tau=u_\tau D \rho/2\mu$, close to the ones  obtained for the slug flow, with values of $Re_{\tau,w} (S_2,S_{3b},S_4, S_{4b}) = [181, \, 161,\, 161, \, 131]$ for the water phase and $Re_{\tau,o} (S_2,S_{3b},S_4, S_{4b}) = [102,\, 91,\, 90,\, 93]$ for the oil phase. 
 The mean profiles of the oil phase, displayed in figure \ref{fig:DNS_slug_mean_flow_oil}, all share similar characteristics. They are all flatter in the bulk than the single phase profiles at similar $Re_\tau$, but in general closer to the single phase profiles than those of the water phase, shown in figure \ref{fig:DNS_slug_mean_flow_wat}. In contrast, the water profiles are extremely flat in the bulk region, with a steep gradient close to the wall. The water profile of $S_4$ is relatively less flat in the bulk than the other cases, closer to those found in the oil phase. The most likely explanation is simply the size of the slug: in $S_4$, the water slug is larger, with a length of around 2.2 diameters, while the other cases had water slugs of approximately $1.1$ or $0.55$ diameters. This influences the calculated mean profile in two different ways. First, in a short slug, the flow far away from one interface might not have enough time to develop before encountering the other interface. Additionally, shorter slugs mean a larger impact from the near-interface region in the simple averaging process used. The fact that the mean profiles from the shortest slug, $S_2$, are noticeably flatter in both cases strengthens this argument. Additionally, when plotted in wall units, as in figure \ref{fig:DNS_slug_mean_flow_oil_yplus} and \ref{fig:DNS_slug_mean_flow_wat_yplus} all profiles collapse in the viscous sublayer ($y^+ < 5$). In this region the flow is dominated by viscous forces and the influence of the interface does not disturb the flow the way it does farther away from the wall. On the other hand, no logarithmic region can be found in the slug profiles, even for the water cases with larger $Re_\tau$, because in that region the flow is not dominated by viscous forces anymore and the influence of the distortions caused by the interface are more noticeable. As seen in figure \ref{fig:DNS_slug_mean_flow_oil}, comparing the single-phase profile with $Re_\tau=90$ (dashed-line) with the oil slugs of cases $S_{3b}$, $S_4$ and $S_{4b}$, the collapse close to the wall does not necessarily take place if adimensionalized in bulk units, even when the value of $Re_\tau$ is comparable. For the collapse to be observed in bulk units, the bulk velocity (adimensionalized with $u_\tau$) has to be similar, which in practice means that the the mean profiles at every radial position  (not just close to the wall) must agree.  In the particular case shown in figure \ref{fig:DNS_slug_mean_flow_oil}, the oil slugs agree fairly well with each other, because they all suffer the same distortions caused by surface tension, and therefore show a good match in both representations. If we were to compare two samples of single-phase flows with similar $Re_\tau$, we would observe a good match in both wall and bulk units as well, because again in that case the dimensionless mean profiles are equal at every point. However, the difference in the bulk profiles between single-phase flow and slug-flow mean that a collapse between the profiles in wall units does not translate into matching curves in bulk units.

\begin{figure} 
 \centering
\begin{subfigure}[c]{0.4\textwidth}
\includegraphics[width=0.90\textwidth]{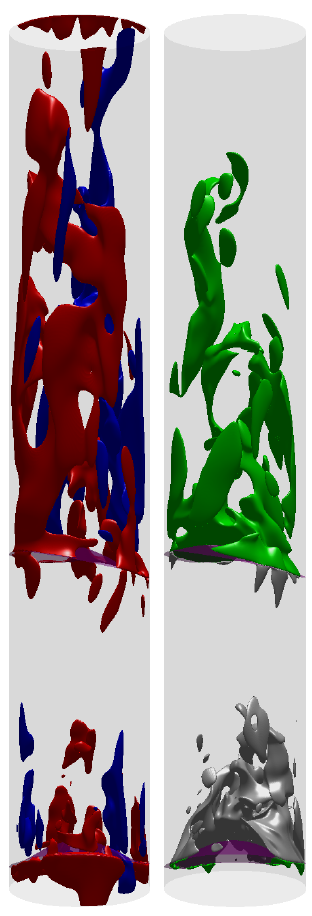}
\subcaption{t = 51.8.0} \label{fig:slug_flow_SLL_KE_VZM_1} 
\end{subfigure} 
\begin{subfigure}[c]{0.4\textwidth}
\includegraphics[width=0.90\textwidth]{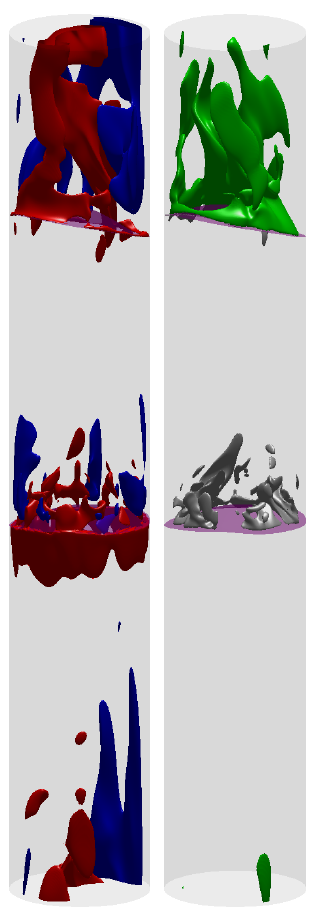}
\subcaption{t = 60.6} \label{fig:slug_flow_SLL_KE_VZM_2} 
\end{subfigure} 

\caption{Isosurfaces of slug flow in $S_{3b}$. Purple: Fluid interface. Red: $u'_{z} = 0.25$. Blue: $u'_{z} = -0.25$. Green/Gray: $E^{r\theta} = 0.019$ (Oil/Water). Time is referenced to the start of its own run.
\label{fig:slug_flow_SLL_KE_VZM}}
\end{figure}

In figure~\ref{fig:slug_flow_SLL_KE_VZM} we show streaks of high (red) and low (blue) axial fluctuation velocity calculated as 
 \begin{equation} \label{eq:fluc_u}
 u'_{z,o/w} = u_z - \bar{u}_{z,o/w},
 \end{equation}
along with isosurfaces of $E^{r\theta}$ for $S_{3b}$. In both cases,  there are large scale structures in each phase, which start at the interface between the slugs and spread downstream from it. As surface tension acts on the surrounding fluid, it generates turbulence that then propagates downstream, but decays in intensity before reaching the other end of the slug. This also forces a large-scale reorganization of the flow, that can be observed in the shape of the mean profile for the two phases, as seen in figure~\ref{fig:DNS_slug_mean_flow}. The large-scale streaks of axial velocity, more intense in the oil slug but also present in the water slug, point to the presence of non-axisymmetric recirculation patterns in the slugs, originated by the interactions at the interfaces.

\subsection{Influence of interface thickness}\label{sec:res_test}

\begin{figure}[!ht]
        \centering
            \centering 
            {{\small }}   
            \includegraphics[scale=0.85]{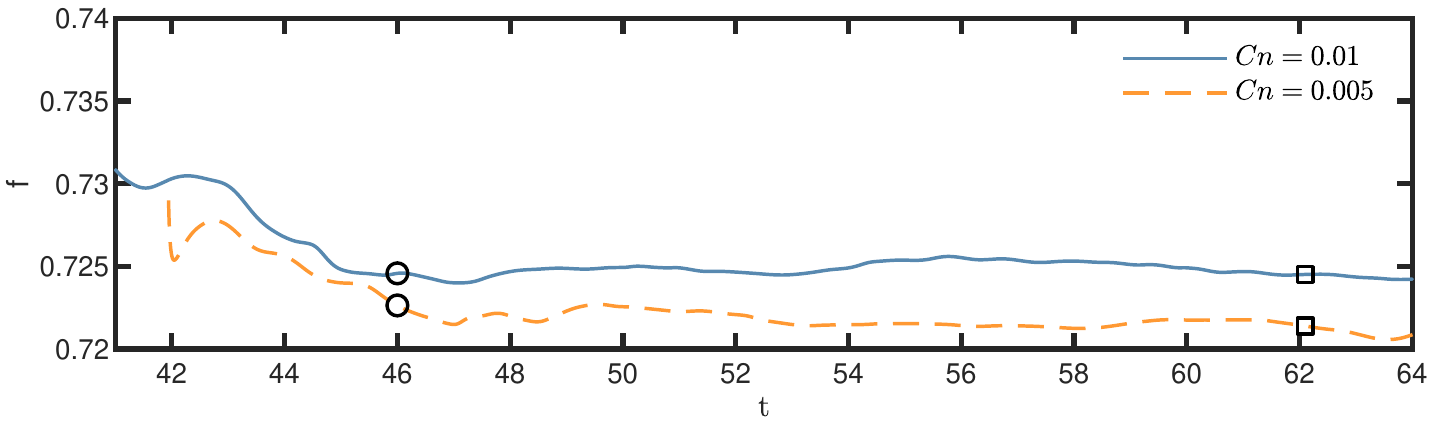}
        \caption[]{\label{fig:Cn_comparison_dpdx}
Influence of $Cn$ number in $S_2$: Temporal evolution of the driving pressure gradient $dp/dx$. Symbols denote the time of the snapshots in figure \ref{fig:Cn_comparison_vec}.}    
\end{figure}

\begin{figure}
\captionsetup[subfigure]{justification=centering}
\begin{center}
\begin{subfigure}[c]{0.20\textwidth}
\includegraphics[width=0.95\textwidth]{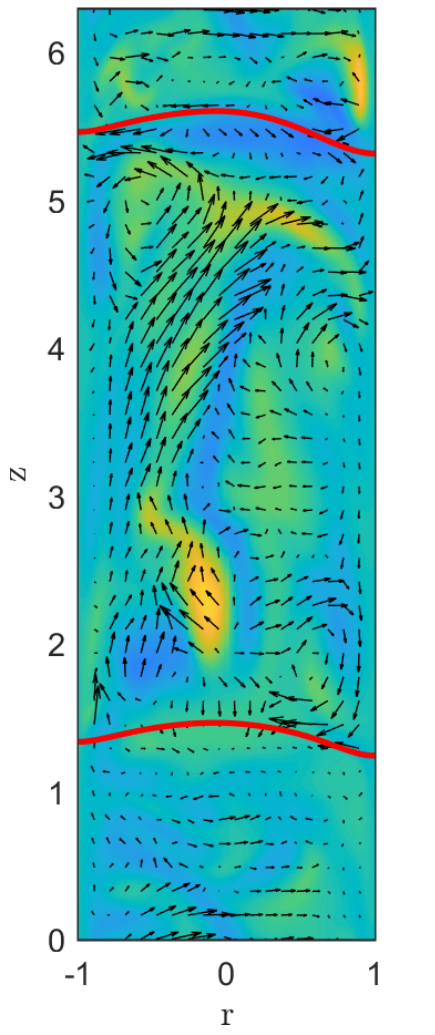}
\subcaption{\footnotesize{t=46.0 \\ Cn=0.01}} \label{fig:Cn_comparison_vec_1} 
\end{subfigure} 
\begin{subfigure}[c]{0.20\textwidth}
\includegraphics[width=0.95\textwidth]{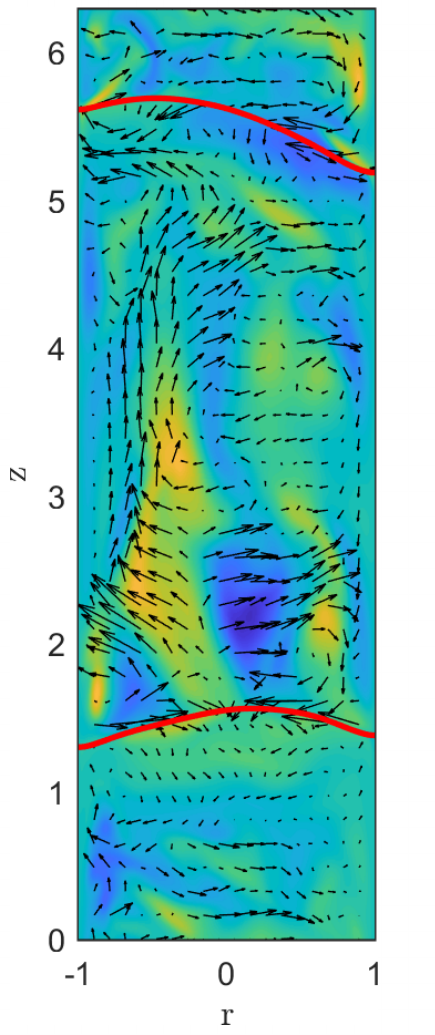}
\subcaption{\footnotesize{t=46.0 \\ Cn=0.005}} \label{fig:Cn_comparison_vec_2} 
\end{subfigure} 
\begin{subfigure}[c]{0.20\textwidth}
\includegraphics[width=0.95\textwidth]{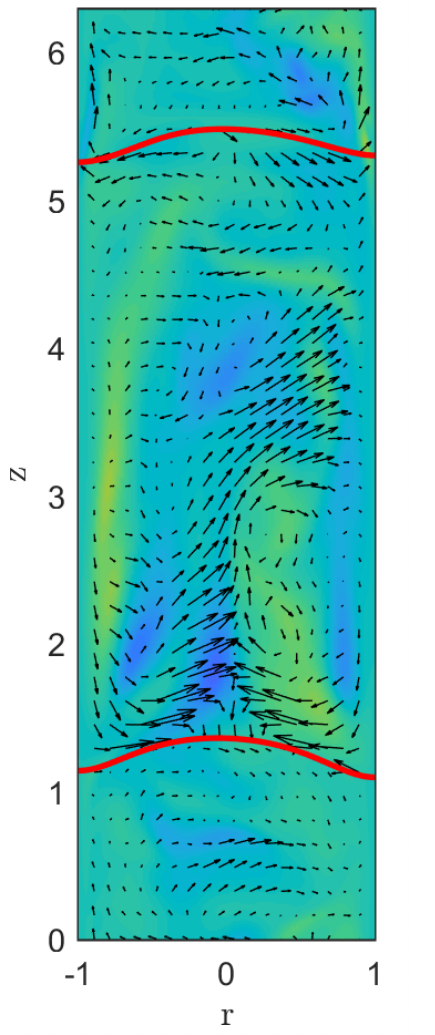}
\subcaption{\footnotesize{t=62.1 \\ Cn=0.01}} \label{fig:Cn_comparison_vec_3} 
\end{subfigure} 
\begin{subfigure}[c]{0.20\textwidth}
\includegraphics[width=0.95\textwidth]{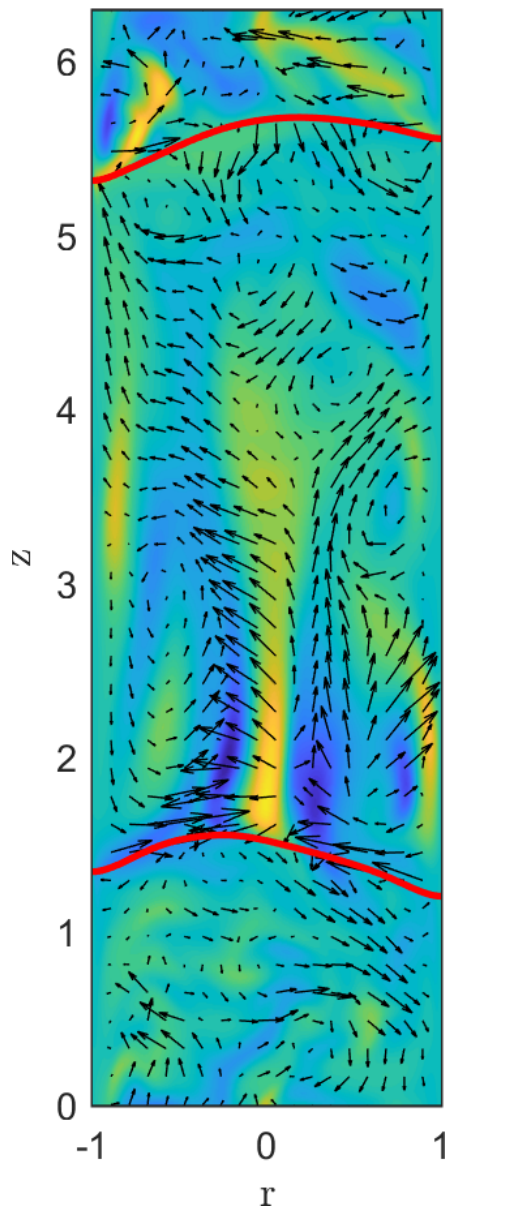}
\subcaption{\footnotesize{t=62.1 \\ Cn=0.005}} \label{fig:Cn_comparison_vec_4} 
\end{subfigure}

\caption{Influence of $Cn$ in $S_2$: Recirculation patterns in the slugs. The red line shows the fluid interface, the Vector field, $(u_r,u_z)$, and the colormap, $u_\theta$. 
 \label{fig:Cn_comparison_vec}}
 
\end{center}
\end{figure}

In the Cahn--Hilliard phase-field method, the interface thickness is set by the dimensionless Cahn number,  $Cn$. To test its influence on the system's evolution and its final state, we performed an additional direct numerical simulation, with $Cn = 0.005$, with half time-step and double resolution in each direction, meaning $\Delta t = 2.5\cdot 10^{-4}$ and $n_r=192$, $2M=512$, $2K=576$. The snapshot of run $S_2$ at $t=42$ was selected as initial condition. At this point in time the final slug is still forming, which allows to test whether the simulation with larger resolution relaxes to the same state or not. In figure~\ref{fig:Cn_comparison_dpdx}, we show the required driving pressure gradient in both cases. After an initial jump, due to the disturbance caused by the relaxation of the interface to the new equilibrium profile, both runs follow the same evolution, with only a small discrepancy between them. Specifically, the run with thinner interface ($Cn=0.005$) presents a smaller pressure gradient required to achieve the same (imposed) average velocity. This is due to its lower dissipation at the interface. Additionally, we show in figure~\ref{fig:Cn_comparison_vec} the fluctuation velocity field (i.e.\ subtracting the mean profile for each phase, see equation \ref{eq:fluc_u}) in the $(r,z)$-plane for the slug flow of $S_2$ with both values of $Cn$. While keeping in mind the limitations of this kind of analysis mentioned in section \ref{sec:analysis_slug}, it helps to confirm the presence of chaotic recirculation structures inside the slugs, and provides evidence that they are correctly captured with $Cn = 0.01$. A difference is that for $Cn=0.005$ the velocity fluctuations appear stronger, consistent again with a lower dissipation at the interface. Overall, we conclude that the results are qualitatively similar and hence that $Cn=0.01$ is sufficient in the simulations to study the regime transitions and properties. 

\subsection{Comparison with experiments}

Direct comparison of simulation with experiments in two-phase flow is challenging. The approach of each method is different, the simulation setting the ratio of volumes and the experiment the ratio of volume flow rates \cite{Bai1992}. Furthermore, the sheer number of dimensionless parameters governing two-phase pipe flow make any comparison difficult, with large differences in the fluid properties \cite{Govier1961, Bai1992}, the pipe orientation \cite{Angeli2000} or simply the pipe diameter \cite{Vigneaux1988}.  Nevertheless, it is still interesting to compare our results with experiments performed under similar conditions. Ghosh \emph{et al.}~\cite{Ghosh2012} worked with a mixture of kerosene and water reporting the flow regime map in the downflow configuration in a pipe of similar diameter, $D =0.012$ cm.  We converted to physical units our superficial velocities in the slug regime ($u_{s,w}=0.13$, $u_{s,o}=0.24$ m/s) and plugged them in their flow regime map for the kerosene-water mixture. Our simulations lay in the boundary between slug flow and stable CAF, which does not contradict our results, even though they worked with a different pipe configuration (downflow vs upflow). The comparatively flat interface and the lack of a water film in our simulation contrasts with the concave interface and water film between the kerosene and the water they reported in the slug regime. This difference is not surprising, since they used hydrophylic walls and we assumed a neutral wall interaction.  

\section{Conclusion and outlook} \label{conclusion}

Phase-field methods have been extensively used in the simulation of multiphase flows, usually with the goal of proving the capabilities of the method to deal with topological changes, and therefore at relatively low $Re$ \cite{Dong2012,Jacqmin1999}. Some authors have reported highly-resolved direct numerical simulations at large values of $Re$ \cite{Soligo2019,Scarbolo2016,Roccon2021} for channel geometries and matched fluid properties, whereas experiments are generally carried out in circular pipes and for fluids of different densities and viscosities. Here we presented  DNS  of an oil-kerosene mixture flowing upward in a pipe with realistic experimental conditions. For this purpose, we employed axially periodic boundary conditions in pipes of up to $4\pi$-diameters in length.  

In the past, linear stability analyses have been shown to produce accurate predictions of the flow patterns for the case of CAF of water and heavy oils at low $Re$ \cite{Bai1992,Joseph1997}. Here we show that a similar approach is not as useful for a water-kerosene mixture. We show that the leading eigenmodes control the dynamics only until the interface touches the wall;  the final saturated state cannot be predicted from the linear stability analysis. The considerable, initial non-modal transient growth of the perturbed modes means that in a situation with a large initial perturbation, such as in experiments, the influence of non-modal interactions might be considerable \cite{Orazzo2014}. Additionally, we showed that selecting the pipe length of the DNS according to the linear stability analysis (most unstable mode) may lead to unphysical results, because the natural structures may not be accommodated therein. While in some cases it might be a reasonable approach when the system tends to converge into a regime closer to the original state, such as bamboo-wave CAF \cite{Song2019}, under the present conditions it results in a non-physical saturated state, specifically the wavy stratified flow shown here. We note that sufficiently long pipes are needed also for correctly capturing the turbulence transition in single-phase pipe flow \cite{Avila2010}.

In sufficiently long pipes, we observed that either a single drop or a slug configuration are found in our simulations. Under realistic experimental conditions, where pipes are much longer, we expect that the flow will initially evolve into a mixture of slugs and drops. As drops are only stable in isolation, a slug following a drop will catch up with it and absorb it, thereby increasing the slug length. Neighboring slugs do not interact because they travel at the same speed. Hence in experiments we expect slugs to dominate sufficiently far away from the pipe entrance~\cite{Ghosh2012}. The flow patterns in the slugs are fairly turbulent and suggest the presence of large scale non-axisymmetric recirculation structures in both phases. Turbulence in the short slugs computed here differs substantially from single-phase turbulence. In much longer slugs ($\gg10D$), fully developed turbulent pipe flow may be expected within. Finally, we stress that the level of turbulence found in the slugs near the interfaces is sensitive to the interface thickness of the phase-field model: thicker interfaces result in increased dissipation and hence smaller velocity fluctuations. Similarly, in the linear stability analysis the enhanced dissipation of a thicker interface renders a smaller growth rate. 

In conclusion, our results demonstrate that phase-field methods are a viable option to numerically explore the flow regime maps of multiphase pipe flows with real fluid mixtures at moderate $Re$.

\section*{Acknowledgement}
The work was supported by the North-German Supercomputing Alliance (HLRN). B.~S. acknowledges financial support from the National Natural Science Foundation of China under grant number 91852105.
\newpage
\section*{References}

\bibliographystyle{ieeetr}
\bibliography{CAF_paper}

\end{document}